\documentclass[conference]{IEEEtran}
\IEEEoverridecommandlockouts

\usepackage{cite}
\usepackage{amsmath,amssymb,amsfonts}
\usepackage{algorithmic}
\usepackage{graphicx}
\usepackage{textcomp}
\usepackage{xcolor}
\usepackage{array}
\usepackage{booktabs}
\usepackage{makecell}
\usepackage{multirow}
\usepackage{enumitem}
\usepackage{comment}
\usepackage{pifont}
\usepackage{tikz}
\usetikzlibrary{arrows.meta,calc,decorations.pathreplacing,fit,positioning,shapes.geometric}
\usepackage[frozencache]{minted}
\usepackage{inconsolata}
\usepackage{listings}
\usepackage[hyphens]{url}
\usepackage{refcount}
\usepackage{cleveref}

\newcommand{\codeunderscore}{\raisebox{0.15ex}{\textunderscore}}
\newcommand{\codedoubleunderscore}{%
  \makebox[1em][c]{\rule[0.08ex]{0.84em}{0.08ex}}%
}
\newcommand{\code}[1]{%
  \begingroup
  \renewcommand{\_}{\codeunderscore}%
  \texttt{#1}%
  \endgroup
}
\newcommand{\myparagraph}[1]{\vspace{0.05in}\noindent\textbf{#1:}}
\newcommand{\etal}{\textit{et al.}}
\renewcommand{\vec}[1]{\boldsymbol{#1}}
\newcommand{\cmark}{\ding{51}}
\newcommand{\xmark}{\ding{55}}

\definecolor{darkblue}{RGB}{0,71,171}

\definecolor{codebg}{gray}{0.95}
\setminted{
  fontsize=\footnotesize,
  bgcolor=codebg,
  linenos,
  numbersep=6pt,
  frame=none,
  breaklines,
}
\def\BibTeX{{\rm B\kern-.05em{\sc i\kern-.025em b}\kern-.08em
    T\kern-.1667em\lower.7ex\hbox{E}\kern-.125emX}}
\newcommand{\IEEEauthorcell}[4]{%
\IEEEauthorblockN{\makebox[0.31\textwidth][c]{#1}}
\IEEEauthorblockA{
\makebox[0.31\textwidth][c]{\textit{#2}}\\
\makebox[0.31\textwidth][c]{#3}\\
\makebox[0.31\textwidth][c]{#4}}}
\usepackage{latexml}
\iflatexml
  \newcommand{\IEEEauthorrowseparator}{\and}
\else
  \newcommand{\IEEEauthorrowseparator}{\and[\hfill\break\null\hfill]}
\fi
\begin{document}

\bstctlcite{flashgpusim:BSTcontrol}

\title{FlashGPU-sim: Enabling GPU Modeling for Modern Architectures and AI Workloads
\thanks{
This work is partially supported by
The Research Grants Council of Hong Kong SAR (No.~T46-415/25-R) and Research Committee of The Chinese University of Hong Kong (Direct Grant No.~518491688).
}
}

\iflatexml
\author{\IEEEauthorcell
{Siying Yu}
{Chinese University of Hong Kong}
{Hong Kong, China}
{syyu25@cse.cuhk.edu.hk}
\and
\IEEEauthorcell
{Yixun Hong}
{Zhejiang University}
{Hangzhou, China}
{foreverhyx@zju.edu.cn}
\and
\IEEEauthorcell
{Guozhi Qiu}
{Zhejiang University}
{Hangzhou, China}
{morettik967@gmail.com}
\and
\IEEEauthorcell
{Jingci Liu}
{Chinese University of Hong Kong}
{Hong Kong, China}
{jcliu@link.cuhk.edu.hk}
\and
\IEEEauthorcell
{Feng Gu}
{Chinese University of Hong Kong}
{Hong Kong, China}
{fgu25@cse.cuhk.edu.hk}
\and
\IEEEauthorcell
{Chenbo Geng}
{Shanghai Jiao Tong University}
{Shanghai, China}
{chenbogeng@sjtu.edu.cn}
\and
\IEEEauthorcell
{Zhengrong Wang\textsuperscript{*}}
{Chinese University of Hong Kong}
{Hong Kong, China}
{zhengrongwang@cuhk.edu.hk}
\and
\IEEEauthorcell
{Chen Zhang}
{Shanghai Jiao Tong University}
{Shanghai, China}
{chenzhang.sjtu@sjtu.edu.cn}
\and
\IEEEauthorcell
{Bei Yu}
{Chinese University of Hong Kong}
{Hong Kong, China}
{byu@cse.cuhk.edu.hk}
\thanks{\textsuperscript{*}Corresponding author.}
}
\else
\author{\IEEEauthorcell
{Siying Yu}
{Chinese University of Hong Kong}
{Hong Kong, China}
{syyu25@cse.cuhk.edu.hk}
\and
\IEEEauthorcell
{Yixun Hong}
{Zhejiang University}
{Hangzhou, China}
{foreverhyx@zju.edu.cn}
\and
\IEEEauthorcell
{Guozhi Qiu}
{Zhejiang University}
{Hangzhou, China}
{morettik967@gmail.com}
\IEEEauthorrowseparator
\IEEEauthorcell
{Jingci Liu}
{Chinese University of Hong Kong}
{Hong Kong, China}
{jcliu@link.cuhk.edu.hk}
\and
\IEEEauthorcell
{Feng Gu}
{Chinese University of Hong Kong}
{Hong Kong, China}
{fgu25@cse.cuhk.edu.hk}
\and
\IEEEauthorcell
{Chenbo Geng}
{Shanghai Jiao Tong University}
{Shanghai, China}
{chenbogeng@sjtu.edu.cn}
\IEEEauthorrowseparator
\IEEEauthorcell
{Zhengrong Wang\textsuperscript{*}}
{Chinese University of Hong Kong}
{Hong Kong, China}
{zhengrongwang@cuhk.edu.hk}
\and
\IEEEauthorcell
{Chen Zhang}
{Shanghai Jiao Tong University}
{Shanghai, China}
{chenzhang.sjtu@sjtu.edu.cn}
\and
\IEEEauthorcell
{Bei Yu}
{Chinese University of Hong Kong}
{Hong Kong, China}
{byu@cse.cuhk.edu.hk}
\thanks{\textsuperscript{*}Corresponding author.}
}
\fi

\iflatexml
\else
\IEEEaftertitletext{\vspace{-2\baselineskip}}
\fi
\maketitle

\begin{abstract}
As AI becomes increasingly ubiquitous, modern AI systems are shaped by a tight software-hardware co-design loop.
Later GPUs expose features such as asynchronous data movement, tensor core pipelines, and fine-grained synchronization that high-performance kernels aggressively exploit, while emerging application behaviors increasingly influence the next generation of hardware design.
Unfortunately, the latest open-source simulators for NVIDIA GPUs
focus on architectures and software stacks from roughly six years ago.
Therefore, they cannot support many state-of-the-art AI kernels generated by
modern compiler stacks, e.g. Triton, or accurately model the hardware features
they depend on. As a result, architects lack a credible platform for analyzing
bottlenecks in this flywheel or evaluating design trade-offs for future AI
systems.

To bridge this gap, we present FlashGPU-sim, an open-source,
execution-driven, cycle-accurate GPU simulator for modern AI workloads.\footnote{\url{https://github.com/FlashGPU-Sim/FlashGPU-Sim}}
FlashGPU-sim faithfully models modern hardware features such as asynchronous data movement, fine-grained synchronization, tensor-core execution, and distributed shared memory.
A Triton extraction front-end allows direct simulation of optimized AI operators without manual porting, while multi-threaded execution makes large-scale software-hardware co-design practical.
Across 131 workload configurations on RTX~5090, H100, and B200, FlashGPU-sim
achieves a cycle-level MAPE of 5.24\%, while multi-threaded simulation
reaches a $7.86\times$ speedup with 16 host threads.
An H100 case study further demonstrates its utility for microarchitectural design exploration.
\end{abstract}

\begin{IEEEkeywords}
GPGPU, GPU Microarchitecture, Modeling, Simulation, Validation, NVIDIA, Hopper, Blackwell.
\end{IEEEkeywords}

\section{Introduction}
\label{sec:intro}

\begin{table*}[t]
\centering
\small
\caption{Comparison of publicly available GPU simulators and FlashGPU-sim. Existing open-source simulators stop at pre-Hopper architectures and omit modern asynchronous GPU execution mechanisms.}
\label{tab:sim-comparison}
\begin{tabular}{lccccccc}
\toprule
\textbf{Simulator} & \textbf{Year} & \textbf{Target Platform} & \textbf{Latest Arch} & \textbf{Arch Release} & \textbf{Method} & \makecell{\textbf{Modern AI}\\\textbf{Workloads}} & \makecell{\textbf{Multi-thread}\\\textbf{Acceleration}} \\
\midrule
GPGPU-Sim~\cite{gpgpusim}                     & 2009 & NVIDIA & Volta   & 2017 & Execution        & \xmark & \xmark \\
Multi2Sim~\cite{multi2sim}                     & 2012 & Both   & Kepler  & 2012 & Execution        & \xmark & \xmark \\
gem5-gpu~\cite{gem5gpu}                        & 2015 & NVIDIA    & Fermi   & 2010 & Execution        & \xmark & \xmark \\
MGPUSim~\cite{mgpusim}                        & 2019 & AMD    & GCN3             & 2016 & Execution        & \xmark & \cmark \\
Accel-Sim~\cite{accelsim}                      & 2020 & NVIDIA & Ampere  & 2020 & Trace \& Execution  & \xmark & \xmark \\
Huerta \etal~\cite{Micro2025DissectingGPUCore} & 2025 & NVIDIA & Ampere  & 2020 & Trace \& Execution  & \xmark & \cmark \\
\midrule
\textbf{FlashGPU-sim}                          & 2026 & NVIDIA & Blackwell & 2024 & Execution     & \cmark & \cmark \\
\bottomrule
\end{tabular}
\end{table*}

Modern AI systems are increasingly shaped by a tight software-hardware co-design loop. GPUs continue to expose new mechanisms for higher-throughput tensor computation and more aggressive overlap between data movement and execution. At the software level, programmers have long relied on hand-written high-performance kernels for the most critical operators, while modern compiler frameworks such as Triton, PyTorch~2, and TileLang are also becoming increasingly important by automatically generating and optimizing kernels to exploit these emerging hardware features~\cite{triton,torch-compile,tilelang}. Large language model serving frameworks further amplify pressure on memory movement, synchronization, and efficiency at scale~\cite{vllm,sglang}. Most importantly, across both hand-written and compiler-generated paths, achieving high performance increasingly depends on how effectively kernels leverage the latest GPU mechanisms. As a result, understanding modern AI performance requires reasoning jointly about evolving hardware capabilities and the software systems and kernels that organize execution around them.

This shift is especially visible in recent NVIDIA architectures. Hopper~\cite{nvidia2022hopper} and Blackwell~\cite{nvidia2025blackwell} expose primitives such as asynchronous bulk data movement, fine-grained hardware synchronization, and increasingly capable tensor-core execution, and modern high-performance kernels rely on these mechanisms to build deeper software pipelines and reduce the coupling between data movement and compute. These mechanisms can certainly be profiled on existing hardware, but profiling only reveals how today's designs behave. It does not answer the more important architecture question: how should future hardware mechanisms be designed, exposed, or balanced once kernel performance increasingly depends on the interaction among asynchronous transfers, synchronization, and tensor-core execution? Answering that question requires more than counters or timelines from current chips; it requires a model that captures these interactions well enough to support architectural what-if analysis.

Such tight interaction between changing software execution and evolving hardware mechanisms has also long been central to computer architecture research. Traditionally, cycle-accurate simulation has been the community's main tool for studying such interactions because it enables detailed performance analysis and design exploration without access to silicon. Within industry, NVIDIA relies on an internal architectural simulator to evaluate large suites of HPC and ML workloads during the design cycle~\cite{NVAS}. In academia, GPGPU-Sim~\cite{gpgpusim} remains the most widely used open-source framework, and later systems such as Accel-Sim~\cite{accelsim} broadened support through trace-driven workflows and updated ISA-level coverage. However, as Table~\ref{tab:sim-comparison} summarizes, publicly available simulators still largely reflect older architectural assumptions and software workflows. They do not faithfully support the asynchronous primitives or modern workload paths that increasingly define AI execution.

This gap directly limits architecture research. Without a simulator that captures current GPU execution mechanisms and modern AI workloads, researchers are forced into an uncomfortable choice: either study outdated architectural models, or rely on measurements from existing hardware alone. Profiling can reveal counters, aggregate timings, and symptoms of performance bottlenecks, but it cannot serve as a manipulable model for architectural exploration. It can tell us how today's chip behaves; it cannot tell us how performance would change if asynchronous data movement were redesigned, if synchronization mechanisms were altered, or if the balance between tensor-core execution and data supply were shifted in a future architecture. In other words, without a credible simulator for today's execution model, it becomes difficult to reason about tomorrow's design trade-offs.

We address this problem with FlashGPU-sim, an execution-driven, cycle-accurate GPU simulator designed for modern AI workloads and modern GPU execution mechanisms. 
FlashGPU-sim extends GPGPU-Sim with models for asynchronous data movement,
fine-grained synchronization, modern Tensor Core execution, and distributed
shared memory, allowing the simulator to capture the execution behaviors that
increasingly shape AI kernel performance.
To ground these models in reality, FlashGPU-sim characterizes key
microarchitectural parameters using targeted hardware microbenchmarks.
To support realistic software stacks, it also includes a Triton-oriented frontend that extracts optimized operators for standalone simulation while preserving a PTX-based path compatible with compiler-generated workloads more broadly. 
Finally, FlashGPU-sim introduces multi-threaded acceleration so that detailed simulation remains practical even for large AI kernels. 
Together, these capabilities provide a more faithful platform for studying bottlenecks in modern kernels and for evaluating future software-hardware design trade-offs on top of a current execution model.

This paper makes the following contributions:

\begin{itemize}
    \item We present FlashGPU-sim, an open-source, execution-driven,
    cycle-accurate GPU simulator that extends GPGPU-Sim with support for
    modern NVIDIA GPU execution mechanisms and AI workloads.

	\item We characterize key timing and resource behaviors of
	modern GPU mechanisms and incorporate the measurements into FlashGPU-sim.

    \item We build a Triton-oriented workload frontend that extracts optimized
    compiler-generated AI operators for standalone simulation, while preserving
    compatibility with PTX-based workloads more broadly.

    \item We validate FlashGPU-sim across RTX~5090, H100, H200, and B200 using
    primitive-level experiments and AI workloads, demonstrate deterministic
    multi-threaded execution, and use an H100 FlashAttention case study 
	for bottleneck analysis and design exploration.
\end{itemize}

\myparagraph{Paper Organization}
Section~\ref{sec:background} motivates why modern AI kernels increasingly 
rely on newer execution mechanisms and introduces their execution structure.
Section~\ref{sec:primitive} presents the hardware behaviors that FlashGPU-sim models.
Section~\ref{sec:simulator} describes the simulator architecture,
extensions, multi-threaded execution, and Triton workload frontend.
Section~\ref{sec:validation} validates FlashGPU-sim across recent NVIDIA
architectures using primitive-level experiments and end-to-end AI workloads,
and evaluates its simulation performance and design-exploration capability.
Section~\ref{sec:related} discusses prior GPU simulators, adjacent AI modeling 
frameworks, and complementary reverse-engineering efforts, and 
Section~\ref{sec:conclusion} concludes the paper.

\section{Background and Motivation}
\label{sec:background}

Modern AI kernel performance is increasingly determined not only by how much
computation is performed, but by how data movement, synchronization, and
tensor-core execution are organized over time. In this section, we use modern
AI kernels to illustrate how newer GPU mechanisms reshape that execution
structure, and why this shift creates a growing gap between real hardware and
current open-source simulators.

\myparagraph{Case Study: FlashAttention on H100}
Attention is one of the most pervasive operators in modern AI
workloads~\cite{attention}. At a high level, it computes a weighted combination
of value vectors using query-key similarity followed by softmax:
\begin{equation}
    \text{Attention}(\vec{Q}, \vec{K}, \vec{V}) = \text{Softmax}(\frac{\vec{Q}\vec{K}^{\mathsf{T}}}{\sqrt{d}})\vec{V}.
\end{equation}
Because attention is both performance-critical and structurally rich in data
movement and tensor computation, it provides a useful case study for
understanding how modern GPU execution mechanisms affect kernel performance.
FlashAttention~\cite{fa} is a family of high-performance attention kernels
that fuse the attention computation through tiling and online softmax,
substantially reducing off-chip memory traffic. More importantly for our
purpose, FlashAttention is not just an algorithmic optimization: its
performance depends heavily on how data movement, synchronization, and
tensor-core computation are scheduled and overlapped within the kernel. It
therefore offers a concrete example of how execution structure, rather than
arithmetic alone, determines AI kernel performance.

\begin{figure}[t!]
\centering
\iflatexml
\includegraphics[width=\textwidth]{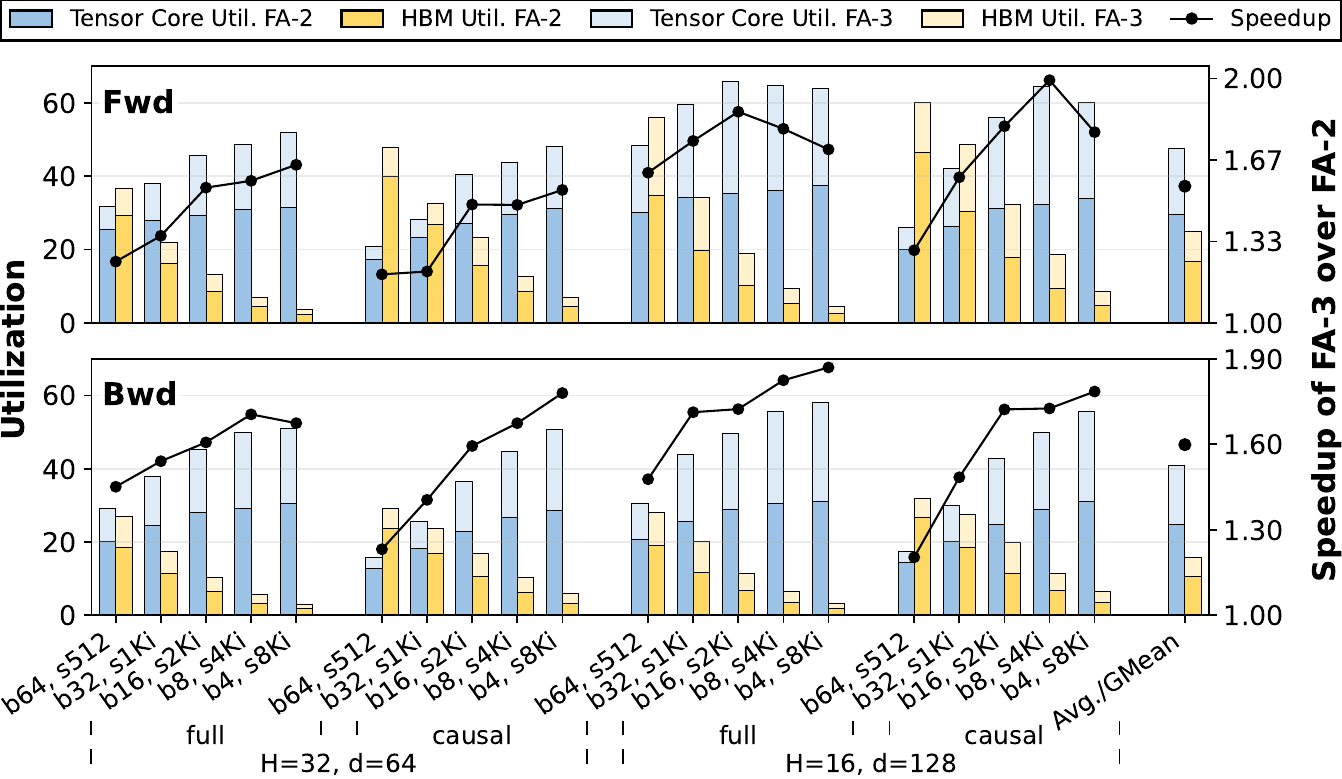}%
\else
\includegraphics[width=\columnwidth]{figs/h100-fa.pdf}%
\fi
\caption{Profiling FlashAttention-3 (FA-3) and FA-2 on H100.}
\label{fig:h100-fa}
\end{figure}

As a concrete example, Figure~\ref{fig:h100-fa} compares
FlashAttention-2~\cite{fa2} and FlashAttention-3~\cite{fa3} on NVIDIA Hopper
across a range of sequence lengths, batch sizes, hidden dimensions, and
forward/backward settings. FlashAttention-3 consistently outperforms
FlashAttention-2, with a geometric-mean speedup of $1.58\times$ and a peak
speedup of $1.99\times$. The key point is not simply that FlashAttention-3
overlaps more operations. Rather, it reorganizes the kernel so that data
movement and pipeline bookkeeping are less tightly coupled to the warps
performing tensor-core computation. As kernels become more compute-dense, this
reduction in resource coupling becomes increasingly important: fewer resources
are tied up in maintaining the pipeline, more execution capacity remains
available for computation, and the kernel can sustain tensor-core throughput
more effectively. To understand why this reorganization matters, we next
examine the execution structures that distinguish these two styles of kernels.

\begin{figure}[t]
\centering
\iflatexml
\includegraphics[width=\textwidth]{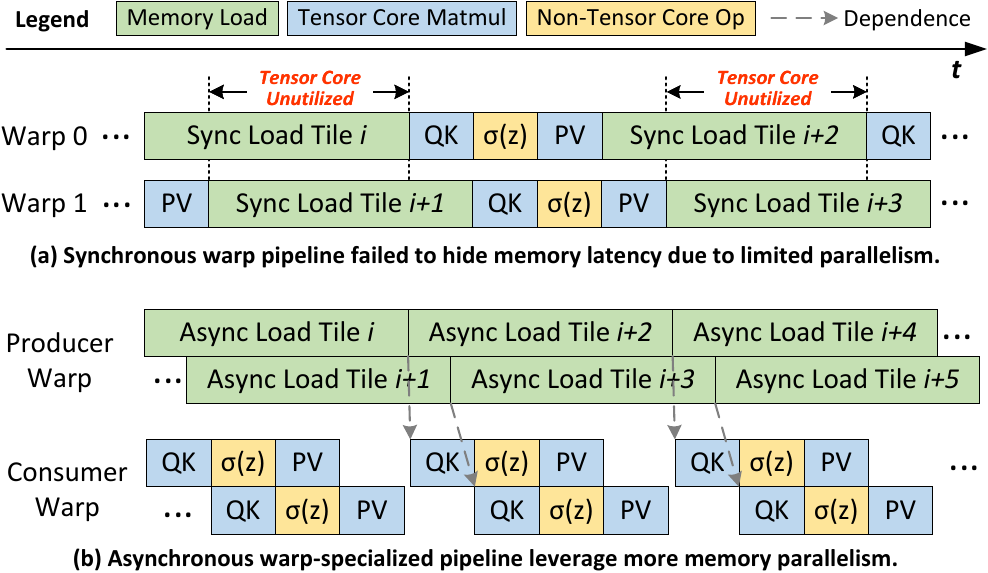}%
\else
\includegraphics[width=\columnwidth]{figs/fa-pipe-crop.pdf}%
\fi
\caption{Synchronous vs Async. Pipeline of Attention.}
\label{fig:fa-pipe}
\end{figure}

\myparagraph{Modern Asynchronous GPU Primitives}
To better understand the performance gap between FlashAttention-2 and
FlashAttention-3, we compare their high-level execution structures in
Figure~\ref{fig:fa-pipe}. Figure~\ref{fig:fa-pipe}(a) illustrates a more
traditional execution style exemplified by FlashAttention-2. All
warps\footnote{In NVIDIA GPU architecture, threads are grouped into warps of 32
threads that execute instructions in lockstep. AMD GPU architectures have
similar concepts with wavefronts of 32 or 64 threads.} move through essentially the
same sequence of work: loading data, performing tensor and non-tensor
computation, and advancing the software pipeline. This homogeneous structure
relies on the GPU's ability to hide latency through massive multithreading, but
it also tightly couples data movement and pipeline bookkeeping to the same
warps that carry out tensor-core computation. That coupling becomes
increasingly costly as the kernel grows more compute-dense. Warps must reserve
more registers and issue slots for tensor-core-heavy computation while still
participating in data movement and control overhead, which reduces how many
warps can remain active at once. With fewer active warps available to absorb
stalls, tensor-core execution becomes easier to starve, and overall performance
degrades.

Hopper addresses this problem by introducing asynchronous primitives that
decouple data movement from computation and enable a more heterogeneous
pipeline structure, as illustrated in Figure~\ref{fig:fa-pipe}(b). A subset of
producer warps can issue asynchronous data movement operations, while consumer
warps concentrate on tensor-core computation without being directly blocked by
those memory operations. Fine-grained synchronization mechanisms enforce the 
ordering between these activities, and warp specialization allows
different warps to assume roles that better match their resource needs. The
importance of these primitives is therefore not limited to accelerating one
specific kernel. Together, they support a different way of organizing
high-performance AI execution, one in which data movement, synchronization, and
computation are less tightly entangled. Once kernel performance increasingly
depends on this organization, the next question is whether current open-source
simulators are able to represent it faithfully.

\myparagraph{Gap between Real Hardware and Current Simulator}
Current open-source simulators are largely unable to faithfully represent the execution of modern AI kernels.
The shift described above is architectural as much as algorithmic. Hopper exposes asynchronous
tensor-memory movement, finer-grained synchronization, and more aggressive
decoupling between data movement and tensor-core execution, while Blackwell
continues this trajectory on newer NVIDIA GPUs~\cite{nvidia2022hopper,nvidia2025blackwell}.
In software, these mechanisms are no longer confined to isolated
microbenchmarks: modern AI kernels increasingly come from both hand-written
implementations and compiler or kernel frameworks such as Triton~\cite{triton},
PyTorch~2~\cite{torch-compile}, TileLang~\cite{tilelang}, and
ThunderKittens~\cite{thunderkittens}, all of which explicitly organize work to
match the memory hierarchy and execution pipeline.

Table~\ref{tab:sim-comparison} makes this modeling disconnect explicit
(detailed comparison in Section~\ref{sec:related}). All prior open-source
simulators in the table omit these asynchronous primitives, and their public
models stop at pre-Hopper architectures. As a result, even when they remain
useful for studying older execution styles, they cannot faithfully represent
modern AI kernels whose performance depends on decoupled data movement,
synchronization, and tensor-core execution. Using such models to reason about
future designs therefore risks drawing conclusions from the wrong execution
model.

\section{Modern GPU Asynchronous Pipeline}
\label{sec:primitive}

\begin{figure}[tb!]
\centering
\iflatexml
\includegraphics[width=\textwidth]{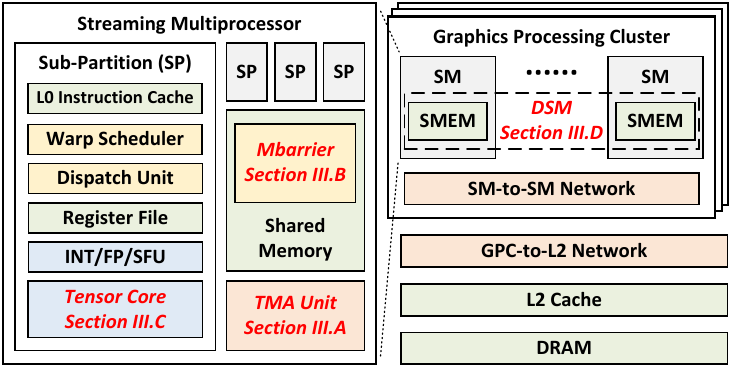}%
\else
\includegraphics[width=\columnwidth]{figs/sm.pdf}%
\fi
\caption{GPU Organization Modeled by FlashGPU-sim.}
\label{fig:sm-overview}
\end{figure}

On recent NVIDIA GPUs, asynchronous execution is realized through three tightly coupled mechanisms: the Tensor Memory Accelerator (TMA) for bulk data movement, the asynchronous memory barrier (mbarrier) for per-stage synchronization, and tensor-core execution for matrix computation, evolving from \code{mma.sync} to \code{wgmma} and further to Blackwell's \code{tcgen05} with Tensor Memory (TMEM).
Figure~\ref{fig:sm-overview} shows how the key components of the asynchronous pipeline are placed within the GPU; the rest of this section presents their microarchitectural models.

FlashGPU-sim extends GPGPU-Sim's existing pipeline abstraction with the components highlighted in Figure~\ref{fig:sm-overview}.
Each streaming multiprocessor (SM) contains four sub-partitions, each with its own warp scheduler, tensor core, and other execution units.
Outside the sub-partitions, a shared TMA unit serves asynchronous data movement requests from all four sub-partitions.
Mbarrier objects reside in shared memory and are operated through shared memory reads and writes; when a TMA transfer completes, the TMA unit signals the associated mbarrier, providing the synchronization link between data movement and computation in the sub-partitions.
At the Graphics Processing Cluster (GPC) level, an SM-to-SM network supports Distributed Shared Memory (DSM), allowing threads in a thread-block cluster to access shared memory (SMEM) across participating SMs.
Together, TMA, mbarrier, and tensor-core execution support the producer-consumer pipeline 
in Figure~\ref{fig:fa-pipe}, while DSM extends data sharing across SMs within a cluster.

NVIDIA documents the instruction set architecture (ISA)-level interface of these mechanisms~\cite{nvidia-ptx-isa} but not their microarchitectural timing characteristics.
We therefore use targeted microbenchmarks to infer the parameters required for cycle-level modeling. 
For this characterization, common mechanisms are studied on RTX~5090. Generation-specific features are characterized on H100 and B200, with H200 additionally used for cluster and DSM characterization.

\subsection{Tensor Memory Accelerator}
\label{subsec:tma}

TMA implements the data-movement stage of the asynchronous pipeline.
On pre-Hopper architectures, loading a tile into shared memory requires threads in the cooperative thread array (CTA) to compute their own addresses and issue their own fine-grained memory instructions.
Ordinary loads additionally route the data through registers before writing it to shared memory; Ampere's \code{cp.async} eliminates this register staging but still requires per-thread address computation and instruction execution, consuming warp scheduler slots and instruction bandwidth for each transfer.

\begin{figure}[t]
\centering
\iflatexml
\includegraphics[width=\textwidth]{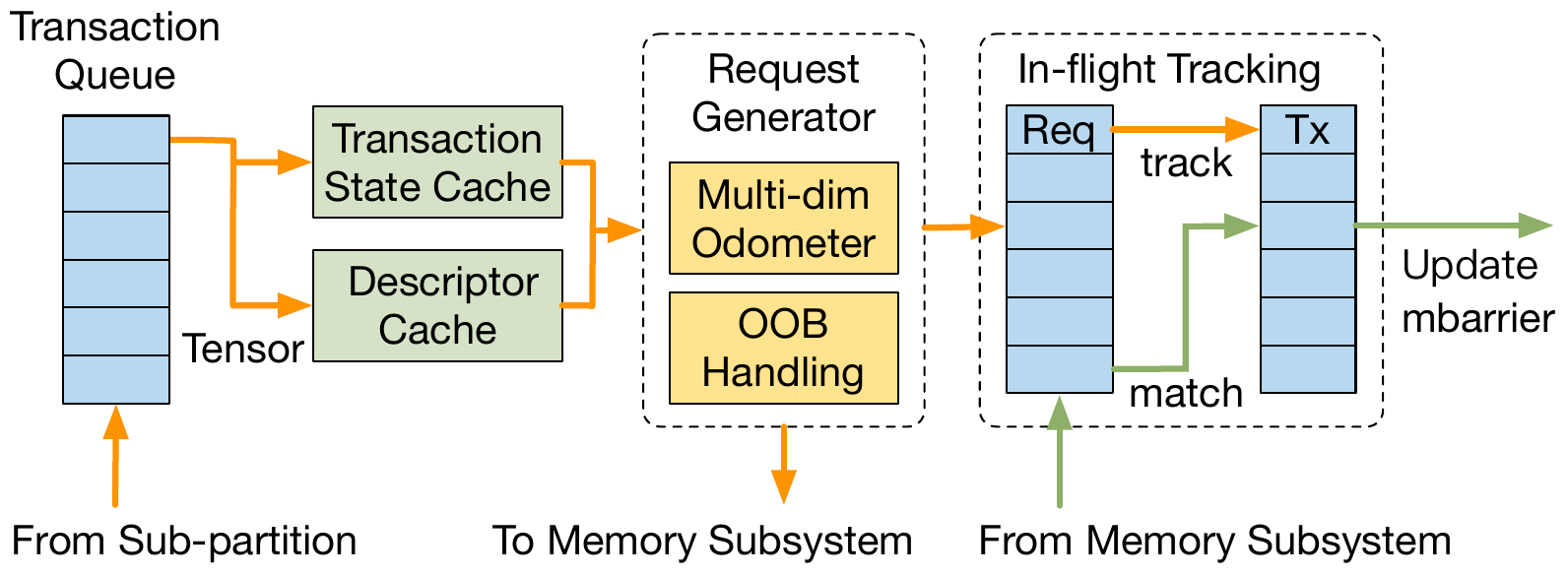}%
\else
\includegraphics[width=1.01\columnwidth]{figs/tma.pdf}%
\fi
\caption{TMA Microarchitecture Model.}
\label{fig:tma-arch}
\end{figure}

Starting with Hopper, the TMA replaces this per-thread pattern with a hardware-managed bulk transfer: a single thread provides a logical coordinate into a multi-dimensional tensor descriptor, while the remaining warps proceed directly with computation.
A dedicated TMA unit on the SM then interprets the tensor descriptor and the instruction's parameters to carry out the transfer: it translates the logical coordinate into physical addresses, generates sector-aligned memory transactions, and writes the result into shared memory, optionally applying swizzling, out-of-bounds clamping, and reduction operations.
NVIDIA's patent filing~\cite{tma-patent} outlines the high-level TMA pipeline. Building on this description together with our microbenchmark measurements, we characterize the microarchitectural model illustrated in Figure~\ref{fig:tma-arch}.

A \code{cp.async.bulk} instruction from a sub-partition is first buffered in the transaction queue.
The TMA unit then initializes the transaction state. In tensor mode, the tensor descriptor is fetched from global memory and cached in a descriptor cache to reduce repeated descriptor accesses.
The request generator iterates over the multi-dimensional address space and generates memory requests while updating the transaction state.
For tensor-mode transfers, it also identifies regions that fall outside the tensor boundary.
Each emitted request is forwarded to the memory subsystem and registered with an internal tracker that associates it with its parent transaction and maintains the number of outstanding requests.
Along the response path, each returning memory response releases its tracking entry, and the tracker detects completion once no outstanding requests remain for the transaction.
The TMA unit then updates the associated \code{mbarrier} to indicate that the transfer is complete.
We next characterize this pipeline's key timing and capacity parameters.

\myparagraph{Transaction Issue}
We first characterize how TMA transactions are issued and how far issue concurrency can extend.
To measure the same-warp TMA issue gap, we use compile-time-unrolled batches of
$N$ \code{cp.async.bulk} instructions with operand preparation and completion waits outside the timed region.
The measured batch execution time $T(N)$ follows $T(N)=1+44N$ cycles exactly across
$N=1$--24, yielding a steady-state issue gap of 44~cycles per TMA instruction.

To determine whether TMA issues across sub-partitions block one another, we compare four transactions under three schedules: \emph{Serial} (issue-and-wait), \emph{Pipelined} (back-to-back issue from one warp), and \emph{Parallel} (concurrent issue from four sub-partitions).
Pipelining reduces completion time from 1669 to 410~cycles, showing substantial transaction overlap.
Parallel issue further reduces the issue span from 248 to 58~cycles, showing little interference among TMA issues from different sub-partitions during concurrent execution.

\begin{figure}[tb!]
\centering
\iflatexml
\includegraphics[width=\textwidth]{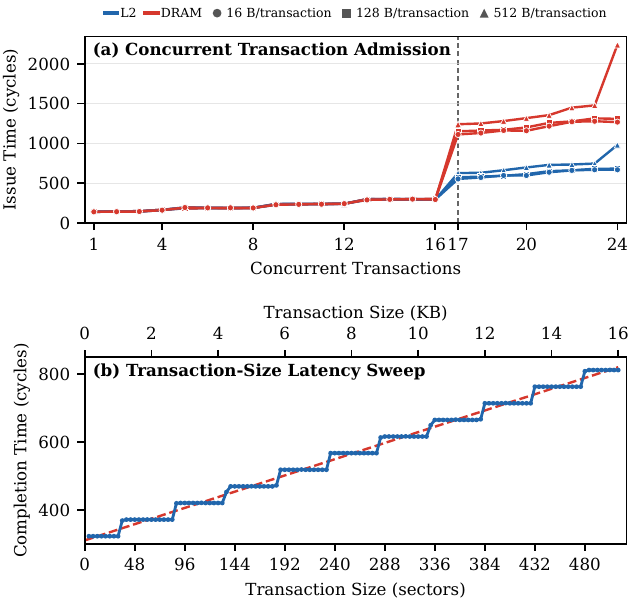}%
\else
\includegraphics[width=\columnwidth]{figs/tma-tx-staircase.pdf}%
\fi
\caption{TMA Transaction Characterization.}
\label{fig:tma-staircase}
\end{figure}

However, this concurrency is bounded by the transaction-admission capacity.
As shown in Figure~\ref{fig:tma-staircase}(a), the issue time rises sharply at the 17th concurrent transaction across different transaction sizes and both L2 and DRAM accesses,
indicating that the TMA front end can admit up to 16 active transactions, with additional issues backpressured.

\myparagraph{Pipeline Throughput}
The TMA pipeline processes memory requests at a fixed rate that bounds the sustained throughput.
NVIDIA Nsight Compute (\code{ncu}) reports a 32~B/cycle per-SM TMA read bandwidth limit.\footnote{Reported by metric \url{l1tex__m_xbar2l1tex_read_bytes_mem_global_op_tma_ld.max.peak_sustained}.}
To validate this limit, we sweep the size of a single TMA transaction from 128~B to 16~KB and measure its completion time, as shown in Figure~\ref{fig:tma-staircase}(b).
The completion time exhibits a clear staircase pattern: completion time remains flat within each plateau and then jumps by approximately 49~cycles every 48~sectors.
Since completion is observed through repeated \code{mbarrier} polling,
the measured latency is quantized by successive barrier checks, whose latency
is characterized in Section~\ref{subsec:mbarrier}.
The staircase slope is $1536\text{B}/49\text{~cycle}\approx32\text{~B/cycle}$, consistent with the \code{ncu}-reported peak.
This limit is architecture specific, reaching 128~B/cycle per SM on B200.

\myparagraph{In-Flight Request Tracking}
Sustaining high TMA throughput relies on maintaining a sufficient number of outstanding memory requests.
By Little's Law, the in-flight tracking capacity can be lower-bounded by the product of the sustained request rate and the access latency.
We therefore evaluate 8~KiB TMA transfers with a 1~GiB working set to expose the DRAM access path.
A single SM sustains 30.816~B/cycle, corresponding to 0.963 requests/cycle at a 32~B granularity.
Combined with a measured DRAM round-trip latency of 845.7~cycles, these observations establish a lower bound of 814 concurrent requests for the underlying tracking capacity.
This confirms that request tracking is adequately provisioned to sustain the observed TMA bandwidth.

\myparagraph{Out-of-Bounds Handling}
Hardware profiling with \code{ncu} reveals asymmetric handling of TMA out-of-bounds (OOB) requests.
For writes, OOB regions generate no additional data traffic.
For reads, OOB accesses are largely prevented from propagating to DRAM, while their handling within the on-chip hierarchy is architecture dependent: RTX~5090 forwards all OOB reads to L2, whereas H100 filters a subset earlier.
To preserve the dominant timing behavior, FlashGPU-sim models OOB reads as reaching L2 without generating downstream DRAM traffic, while matching the architecture-specific filtering details provides negligible benefit to simulation accuracy.
\subsection{Asynchronous Memory Barrier}
\label{subsec:mbarrier}

\iflatexml
\begin{table}[t]
\centering
\caption{Mbarrier polling latency on RTX~5090.}
\label{tab:mbarrier-wait}
\setlength{\tabcolsep}{4pt}
\footnotesize
\begin{tabular}{lll}
\toprule
\textbf{Wait Form} & \textbf{SASS Pattern$^{*}$} & \textbf{Latency} \\
\midrule
\code{test\_wait}
& \code{PHASECHK}
& 42 cy \\

\code{try\_wait}
& \code{PHASECHK.TRYWAIT}
& 42 cy \\

\quad + hint, complete
& \code{TRYWAIT + SLEEP + PHASECHK}
& 150 cy \\

\quad + hint, incomplete
& \code{TRYWAIT + SLEEP + PHASECHK}
& hint dependent \\
\bottomrule
\end{tabular}

\vspace{2pt}
\footnotesize
$^{*}$SASS mnemonics are abbreviated for readability:
\code{PHASECHK} denotes \code{SYNCS.PHASECHK.TRANS64}, and
\code{SLEEP} denotes \code{NANOSLEEP.SYNCS}.
\end{table}

\fi

The \code{mbarrier} primitive provides the fine-grained synchronization needed by modern asynchronous pipelines. 
Because TMA transfers proceed independently of the issuing thread, software-pipelined kernels need a way to determine when asynchronously produced data has become ready for consumption. 
Conventional primitives such as \code{\codedoubleunderscore syncthreads} and \code{\codedoubleunderscore syncwarp} synchronize participating threads at CTA or warp scope, but they do not directly represent the completion of work performed by asynchronous engines.

An mbarrier provides this coordination through a synchronization object residing in shared memory. 
Its state can be updated both by threads through barrier arrival operations and by asynchronous engines such as TMA upon transaction completion. 
Each mbarrier tracks the progress of a synchronization phase, including participating thread arrivals 
and pending transaction bytes, allowing consumers to determine when that phase has completed. 
By assigning separate mbarrier instances to different pipeline stages, producers and consumers can advance on different tiles concurrently, enabling warp specialization and fine-grained overlap among data movement, synchronization, and computation.
This coordination relies on timely phase-completion observation.
We therefore characterize both the latency of individual barrier checks and the
visibility granularity introduced by repeated polling.

\myparagraph{Polling Latency}
\code{mbarrier} supports two ways of observing phase completion:
\code{test\_wait} non-blockingly probes the phase, whereas
\code{try\_wait} may wait for completion and optionally takes a suspend-time
hint.
Table~\ref{tab:mbarrier-wait} summarizes their SASS paths and predicate-ready latencies.
With the barrier state held fixed, \code{test\_wait} and no-hint
\code{try\_wait} map to distinct SASS patterns, yet both produce a valid
predicate after 42 cycles, independent of whether the barrier is complete.

A suspend-time hint selects a longer wait path with an added sleep stage.
For a complete barrier, this path has a 150-cycle result-ready latency
regardless of the hint value, even when set to zero.
When the barrier remains incomplete, the hint governs the maximum suspension time, which scales in discrete steps.
Specifically on RTX~5090, suspension durations are quantized at power-of-two boundaries, 
while hint values within the same interval exhibit similar return-time behavior.

\myparagraph{Completion Visibility}
Polling latency sets the temporal granularity of barrier visibility.
Because a barrier phase may complete between successive polls, the consumer
observes the completion transition only at a subsequent barrier check,
quantizing the observed completion time into discrete steps.
The preceding TMA characterization uses no-hint \code{try\_wait} polling,
producing the observed staircase in completion time.

\iflatexml
\else
\begin{table}[t]
\centering
\caption{Mbarrier polling latency on RTX~5090.}
\label{tab:mbarrier-wait}
\setlength{\tabcolsep}{4pt}
\footnotesize
\begin{tabular}{lll}
\toprule
\textbf{Wait Form} & \textbf{SASS Pattern$^{*}$} & \textbf{Latency} \\
\midrule
\code{test\_wait}
& \code{PHASECHK}
& 42 cy \\

\code{try\_wait}
& \code{PHASECHK.TRYWAIT}
& 42 cy \\

\quad + hint, complete
& \code{TRYWAIT + SLEEP + PHASECHK}
& 150 cy \\

\quad + hint, incomplete
& \code{TRYWAIT + SLEEP + PHASECHK}
& hint dependent \\
\bottomrule
\end{tabular}

\vspace{2pt}
\footnotesize
$^{*}$SASS mnemonics are abbreviated for readability:
\code{PHASECHK} denotes \code{SYNCS.PHASECHK.TRANS64}, and
\code{SLEEP} denotes \code{NANOSLEEP.SYNCS}.
\end{table}

\fi
\subsection{The Evolving Tensor Core}
\label{subsec:tensorcore}

\begin{table}[t]
\centering
\caption{MMA issue gap and execution latency on RTX~5090.}
\label{tab:mma-timing}
\setlength{\tabcolsep}{3pt}
\footnotesize
\begin{tabular}{llccccc}
\toprule
\textbf{Type} & \textbf{Shape} & \textbf{Peak}
& \textbf{Gap$^{\text{theo}}_\text{warp}$}
& \textbf{Gap$^{\text{exp}}_\text{warp}$}
& \textbf{Eff.} & \textbf{Latency} \\
\midrule
\multirow{2}{*}{FP16, FP32}
  & M16N8K16 & 209.5 & 32 & 33.4 & 95.7\% & 34.5 \\
  & M16N8K8  & 209.5 & 16 & 32.8 & 48.9\% & 34.5 \\
\midrule
BF16, FP32
  & M16N8K8  & 209.5 & 16 & 32.8 & 48.9\% & 34.4 \\
\midrule
\multirow{2}{*}{TF32, FP32}
  & M16N8K8  & 104.8 & 32 & 33.4 & 95.7\% & 34.4 \\
  & M16N8K4  & 104.8 & 16 & 32.8 & 48.9\% & 34.5 \\
\midrule
\multirow{2}{*}{INT8, INT32}
  & M16N8K32 & 838.0 & 16 & 19.6 & 81.6\% & 27.0 \\
  & M16N8K16 & 838.0 &  8 & 19.0 & 42.1\% & 27.0 \\
\bottomrule
\end{tabular}

\vspace{2pt}
\footnotesize
Peak is in TOPS at 2407\,MHz; issue gaps and latency are in cycles.
\end{table}

Tensor Core instructions implement the compute stage of the asynchronous pipeline and have evolved across recent GPU generations.
We first characterize \code{mma.sync}, which remains supported across recent architectures, and then examine the generation-specific asynchronous paths exposed through Hopper's \code{wgmma} and Blackwell's \code{tcgen05} with TMEM.
Beyond execution timing, we also recover the underlying arithmetic structure for accurate functional simulation.

\myparagraph{MMA Execution}
The conventional Tensor Core interface is the warp-level matrix multiply-accumulate 
instruction \code{mma.sync}, in which each warp synchronously operates on register fragments.
For \code{mma.sync}, tensor-core performance is shaped by
both the minimum interval between independent instruction issues and the
latency exposed by dependent operations.
We characterize these effects as the issue gap and
execution latency, respectively.

Published RTX~5090 specifications provide a theoretical lower bound on the
\code{mma.sync} issue gap~\cite{nvidia2025blackwell}.
Let $T_\text{chip}$ denote the peak dense tensor throughput in TOPS and $S$ the number of SMs.
An \code{mma.sync} of shape $M{\times}N{\times}K$ performs $2MNK$ operations,
giving the theoretical per-SM issue gap:
\begin{equation}
\label{eq:gap-sm}
\text{Gap}^{\text{theo}}_{\text{SM}} = \frac{2MNK \cdot S \cdot f}{T_\text{chip} \times 10^{3}} \;\text{(cycles)}
\end{equation}
where $f$ is the SM clock frequency in GHz.
Since each SM contains $P{=}4$ tensor-core sub-partitions, the corresponding
single-warp theoretical gap is
$\text{Gap}^{\text{theo}}_{\text{warp}}
= P \cdot \text{Gap}^{\text{theo}}_{\text{SM}}$.

To measure the issue gap and execution latency, we execute $k$
independent \code{mma} chains within a single warp while sweeping the chain length $N$.
Linear regression over $N$ extracts the corresponding slope.
With sufficiently many independent chains, dependencies are hidden and the
slope converges to the saturated issue gap.
With a single dependent chain, each instruction consumes the accumulator
produced by its predecessor, and the slope captures the execution latency.

As shown in Table~\ref{tab:mma-timing}, the measured issue gap and execution latency 
remain approximately constant across instruction shapes within each data type.
The larger shapes approach the theoretical Tensor Core throughput, 
whereas the smaller shapes fall short because they perform fewer operations at the same issue rate.
Section~\ref{subsec:primitive-alignment} further shows that these
timing parameters reproduce the corresponding full-chip throughput.

However, \code{mma.sync} can no longer fully utilize the growing 
Tensor Core capacity of recent datacenter GPUs~\cite{colfax2026blackwellmma}.
For dense FP16 inputs with FP32 accumulation, we measure the saturated throughput of 
\code{mma.sync} to be approximately 66\% of the theoretical per-cycle throughput on H100 and 25\% on B200.
Sustaining full Tensor Core throughput therefore increasingly depends on the asynchronous tensor operations.

\myparagraph{Asynchronous Tensor Operations}
Hopper introduces asynchronous warp-group matrix multiply-accumulate instructions, 
\code{wgmma}, which coordinate tensor operations across four warps.
Blackwell further introduces the fifth-generation Tensor Core instructions, 
\code{tcgen05}, which extend tensor operations to CTA pairs 
and use TMEM for accumulator and operand storage.

\code{wgmma} and \code{tcgen05} share a common compute throughput model.
For each operation, its workload of $2MNK$ FLOPs is converted into a compute
service time using a data-type-specific per-SM service rate.
Based on our measurements, the FP16 service rate is set to
4096~FLOP/SM/cycle on H100 and 8192~FLOP/SM/cycle on B200.
All concurrent tensor operations on the same SM share this service capacity, bounding
their aggregate compute throughput by the configured rate.

However, the compute service rate alone does not capture interference between
\code{wgmma} and concurrent non-tensor computation.
On H100, we observe that softmax execution slows when overlapped with background
\code{wgmma} operations, suggesting contention from the register-file traffic
generated by \code{wgmma}.
To model this effect, each \code{wgmma} operation emits register-traffic proportional 
to its accumulator size, which consumes the register-bandwidth budget shared with the operand collector.
This interference also highlights the architectural rationale for TMEM on Blackwell:
whereas \code{wgmma} keeps accumulators in registers, \code{tcgen05} moves
them to TMEM, reducing register pressure during asynchronous tensor execution.

\myparagraph{Bit-Exact Functional Arithmetic}
To achieve functional fidelity, we further characterize the internal
floating-point arithmetic used by Tensor Core operations.
For finite values, the hardware derives a type-dependent internal least significant bit (LSB)
from the product and accumulator ranges, then truncates and aligns all terms
before accumulation at a shared internal precision.
For non-finite inputs, fixed-priority logic handles NaNs, invalid products,
and infinity-sign conflicts, producing canonical NaNs or propagating valid
infinities.
The resulting arithmetic behavior aligns with the bit-accurate Tensor Core
arithmetic model reported in MMA-Sim~\cite{mma-sim}.
Our implementation covers seven \code{mma.sync} variants with FP32 accumulation 
spanning FP16, BF16, TF32, and E4M3/E5M2 FP8 formats.
Validation with instruction-level probes and GEMM workload replay on an RTX~5090 confirms
that simulated outputs are bitwise identical to hardware outputs.

\iflatexml
\else
\begin{table*}[t]
\centering
\caption{Overview of FlashGPU-sim extensions.}
\label{tab:sim-extension-overview}
\small
\renewcommand{\arraystretch}{1.12}
\setlength{\tabcolsep}{5pt}
\begin{tabular}{@{}lll@{}}
\toprule
\textbf{Category} & \textbf{Extension} & \textbf{Description} \\
\midrule

\multirow{4}{*}{Architectural Support}
& Data Movement
& \code{cp.async}, TMA bulk transfers \\
& Synchronization
& \code{mbarrier}, bulk-group completion \\
& Tensor Core Computation
& \code{mma.sync}, \code{wgmma}, \code{tcgen05}, and TMEM \\
& DSM and Clusters
& Cluster launch, distributed shared memory, and intra-GPC communication \\

\midrule
\multirow{2}{*}{Calibration and Fidelity}
& PTX Reordering
& Instruction reordering, instruction fusion, and redundant instruction elimination \\
& Memory Subsystem
& Memory hierarchy mapping, interconnect modeling, address hashing \\

\midrule
Simulation Performance
& Multi-threaded Acceleration
& Per-SM and per-GPC parallelism, deterministic memory arbitration \\

\midrule
Ecosystem Support
& Triton Frontend
& Kernel and argument capture, harness generation, online and offline replay \\

\bottomrule
\end{tabular}
\end{table*}

\fi

\subsection{Distributed Shared Memory}
\label{subsec:cluster}

Thread block clusters further extend producer-consumer pipelines across CTAs by allowing 
groups of CTAs to co-reside within the same GPC~\cite{nvidia2022hopper,hopper-tuning}.
This enables data movement, synchronization, and computation to overlap across multiple SMs.
Cluster-level communication includes DSM remote accesses, 
shared-to-shared TMA transfers and TMA multicast selected by CTA masks, 
and cross-CTA \code{mbarrier} synchronization~\cite{nvidia-ptx-isa}.
Together, these mechanisms support clustered GEMM and attention kernels.

\begin{figure}[t]
\centering
\iflatexml
\includegraphics[width=\textwidth]{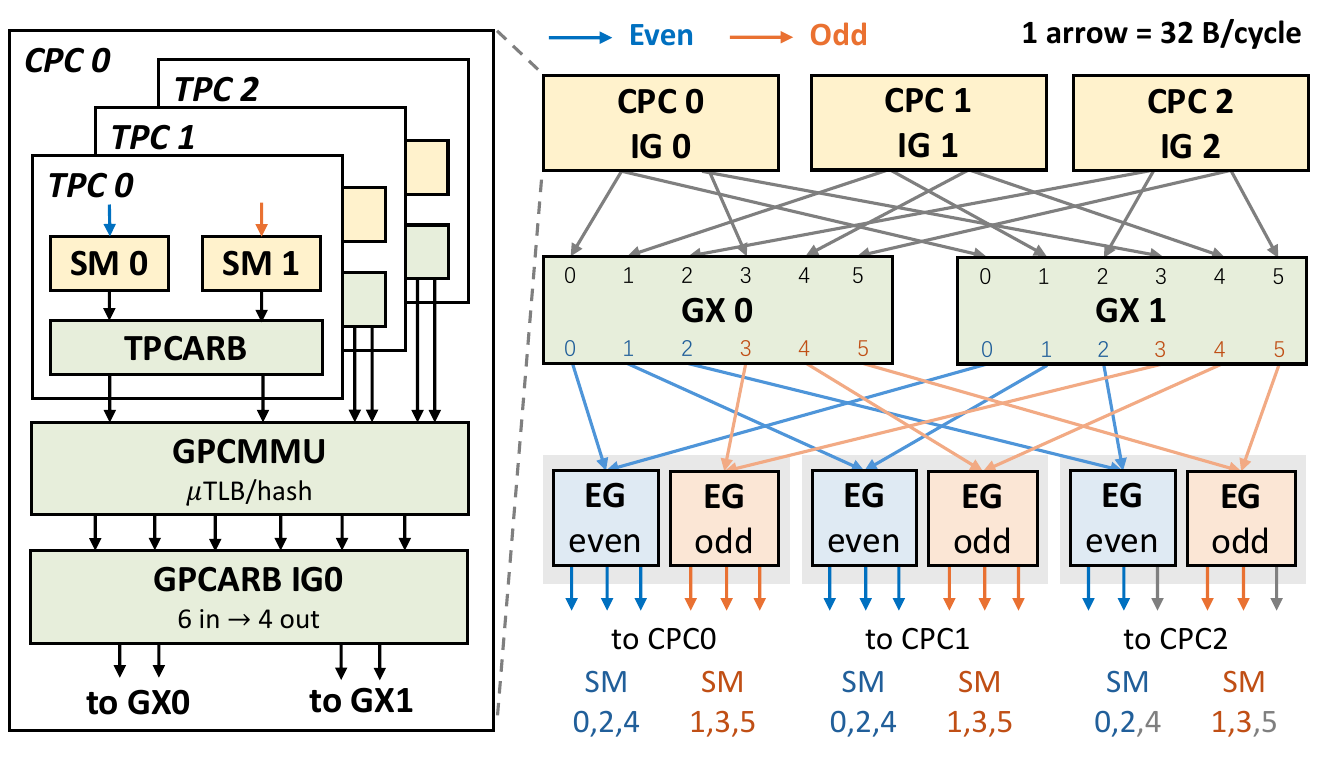}%
\else
\includegraphics[width=\columnwidth]{figs/dsm.pdf}%
\fi
\caption{Intra-GPC DSM Fabric~\cite{us12248788}.}
\label{fig:dsm-fabric}
\end{figure}

\myparagraph{Intra-GPC Fabric}
Supporting these operations requires SM-to-SM communication within the GPC.
Informed by NVIDIA's DSM patent~\cite{us12248788},
FlashGPU-sim models the intra-GPC fabric as shown in Figure~\ref{fig:dsm-fabric}.
Each GPC comprises several Compute Processing Clusters (CPCs). 
Each CPC contains three Texture Processing Clusters (TPCs), with two SMs and one TPCARB per TPC, 
and shares one GPCMMU and one GPCARB.\footnote{TPCARB denotes the TPC arbiter, GPCMMU the GPC memory management unit,
GPCARB the GPC arbiter, and IG/EG its ingress/egress blocks.}
GX0 and GX1 are parallel switch planes; firmware may disable one plane and retain connectivity at reduced bandwidth.

On this topology, DSM loads and stores, remote \code{mbarrier} operations, and shared-to-shared TMA transfers traverse the fabric without occupying the L2 cache.
Remote \code{mbarrier} operations reuse the barrier model of Section~\ref{subsec:mbarrier}: 
a peer CTA sends a short fabric message to the owner SM, which applies the update to the local barrier state.
Separately, global-to-shared TMA multicast bypasses the DSM fabric and is modeled as functional payload fan-out plus a completion delay.

While topology defines connectivity, we further characterize one-way DSM traffic to determine the fabric's per-SM service rate.
Remote loads, stores, and TMA puts on H200 all sustain about 20--21\,B/cycle per SM~\cite{wang2026dsm}.
With each CPC connecting six SMs to four links into the GX switches, 
a 32~B/cycle service rate per link yields $4\times32\,\text{B}/6 \approx 21.3\,\text{B/SM/cycle}$, 
consistent with the measured throughput.
We also observe that a single SM remains near this rate when neighboring SMs are idle, indicating that unused capacity is not redistributed.
FlashGPU-sim therefore caps each SM at two-thirds of one link's service rate, allowing one 32\,B payload in two of every three cycles without reallocating idle slots to neighboring SMs.

However, concurrent DSM traffic exhibits contention.
When two SMs issue remote accesses to each other simultaneously, per-direction read throughput drops by 23\%, whereas store and TMA-put throughput drops by only 4--5\%.
This difference comes from reverse-path traffic: a remote load is modeled as four 32\,B reply payloads plus one reverse read request, while stores and TMA puts send payload forward and return a coalesced acknowledgment completing several writes at once.
In addition, two DSM streams sharing the same direction saturate near 21\,B/cycle, whereas opposite-direction streams achieve 36--39\% higher throughput~\cite{wang2026dsm}.
This suggests a per-SM sending limit: same-direction streams share one SM's budget, while opposite-direction streams use two.
FlashGPU-sim captures this contention with a per-SM send budget shared by request and reply queues.

\section{FlashGPU-sim}
\label{sec:simulator}

\iflatexml
\else
\begin{figure*}[t]
\centering
\iflatexml
\includegraphics[width=\textwidth]{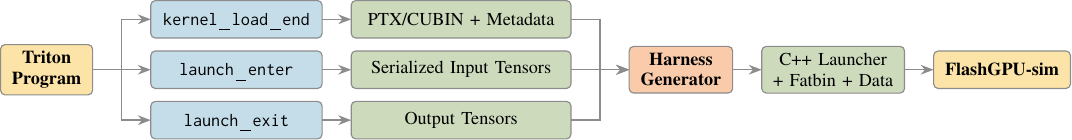}%
\else
\includegraphics{figs/triton_workflow.pdf}%
\fi
\caption{Online Triton Kernel Extraction Workflow.}
\label{fig:triton-workflow}
\vspace{-8pt}
\end{figure*}

\fi

\subsection{System Overview}
\label{subsec:sim-overview}

FlashGPU-sim extends GPGPU-Sim~\cite{gpgpusim} with the modern GPU mechanisms 
characterized in Section~\ref{sec:primitive}. 
The characterized microarchitectural parameters are exposed as configuration entries 
to support cross-architecture calibration and design exploration.
Beyond the architectural support, FlashGPU-sim also broadens ecosystem support, 
improves modeling fidelity and simulation throughput, 
as summarized in Table~\ref{tab:sim-extension-overview}.

\iflatexml
\begin{table*}[t]
\centering
\caption{Overview of FlashGPU-sim extensions.}
\label{tab:sim-extension-overview}
\small
\renewcommand{\arraystretch}{1.12}
\setlength{\tabcolsep}{5pt}
\begin{tabular}{@{}lll@{}}
\toprule
\textbf{Category} & \textbf{Extension} & \textbf{Description} \\
\midrule

\multirow{4}{*}{Architectural Support}
& Data Movement
& \code{cp.async}, TMA bulk transfers \\
& Synchronization
& \code{mbarrier}, bulk-group completion \\
& Tensor Core Computation
& \code{mma.sync}, \code{wgmma}, \code{tcgen05}, and TMEM \\
& DSM and Clusters
& Cluster launch, distributed shared memory, and intra-GPC communication \\

\midrule
\multirow{2}{*}{Calibration and Fidelity}
& PTX Reordering
& Instruction reordering, instruction fusion, and redundant instruction elimination \\
& Memory Subsystem
& Memory hierarchy mapping, interconnect modeling, address hashing \\

\midrule
Simulation Performance
& Multi-threaded Acceleration
& Per-SM and per-GPC parallelism, deterministic memory arbitration \\

\midrule
Ecosystem Support
& Triton Frontend
& Kernel and argument capture, harness generation, online and offline replay \\

\bottomrule
\end{tabular}
\end{table*}

\fi

\myparagraph{Modern Programming Interface}
FlashGPU-sim extends the PTX execution engine and CUDA launch interface to
natively intercept and execute modern GPU workloads.
The simulator interprets execution metadata and launch parameters, such as TMA
tensor descriptors and thread-block cluster configurations, to determine the corresponding runtime behavior.
For example, cluster configuration directs the scheduler to co-locate CTAs from
the same cluster within a single GPC while exposing cluster IDs
and CTA ranks to the executing kernel.

\myparagraph{PTX Reordering}
FlashGPU-sim executes PTX instructions, whereas real GPUs execute SASS code
after backend compiler optimization and scheduling.
Prior work has shown that differences between intermediate and machine-level
instruction streams can substantially affect GPU performance modeling~\cite{gutierrez2018lost}.
These differences arise in multiple forms.
For instance, backend compilation may fuse or eliminate instructions, resulting
in a more concise SASS stream, or reorder instructions to better exploit
instruction-level parallelism.

In particular, our experiments show that instruction reordering has a much
larger impact on performance prediction than instruction-count reduction,
especially for kernels with limited warp-level latency hiding.
To reduce this mismatch, FlashGPU-sim performs static PTX instruction
reordering before simulation.
It constructs a dependency graph and schedules ready
instructions using estimated latencies and pipeline availability, exposing
instruction-level overlap that the original PTX order leaves unexploited.

\myparagraph{Memory Subsystem}
To reproduce the hardware's traffic distribution across memory sub-partitions,
we calibrate the IPOLY hash function~\cite{ipoly} by shifting the hash input
toward lower address bits that capture stride-dependent variation.
Memory service rates also differ across architectures and along the memory path.
For example, the SM-to-crossbar port services 32~B/cycle on RTX~5090
and 128~B/cycle on B200.
To capture these service rates, FlashGPU-sim models independent service limits for
SM-side request and response handling, interconnect transport, L2 access,
and DRAM service.

\myparagraph{Multi-threaded Acceleration}
Cycle-level simulation of large AI kernels can take hours when single-threaded, making multi-threaded acceleration essential.
In the standard execution, each SM maintains independent pipeline state, so we
parallelize the per-SM core cycle across OpenMP threads.
When thread-block clusters are active, DSM fabric state is shared by the SMs within a GPC,
shifting parallel granularity to the GPC level, with one host thread owning each GPC.
Because the interpreter carries per-invocation state in shared instruction objects, we maintain per-thread copies of decoded instructions to ensure concurrent SMs can safely execute the same static instruction.
Detailed speedup measurements are reported in Section~\ref{subsec:simulation-performance}.

To maintain cycle-level determinism without complex synchronization overhead,
we decouple the simulation: SMs or GPCs execute concurrently; the interconnect
and memory subsystems advance sequentially; in cluster mode, the DSM fabric advances
once per GPC per cycle.
During this serial phase, memory requests are buffered into per-source queues, and deterministic crossbar arbitration resolves cross-SM contention, ensuring outcomes are host-scheduling independent.

\subsection{AI Workload Support}
\label{subsec:triton}

\iflatexml
\begin{figure*}[t]
\centering
\iflatexml
\includegraphics[width=\textwidth]{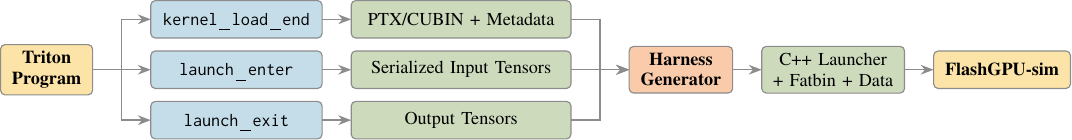}%
\else
\includegraphics{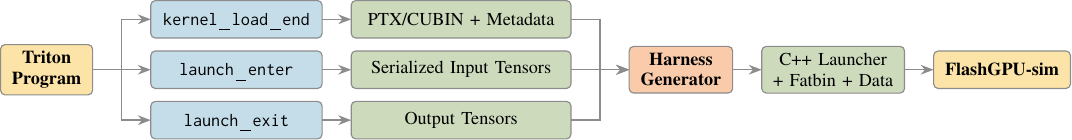}%
\fi
\caption{Online Triton Kernel Extraction Workflow.}
\label{fig:triton-workflow}
\vspace{-8pt}
\end{figure*}

\fi

\begin{figure}[t]
\begin{minted}{python}
import TritonTrace

def extract(mode="online"):
    if mode == "online":
        tracker = TritonTrace.Tracker(
            output_dir, mode="online", enabled=False)
        my_kernel[grid](A, B, C, M, N, K)  # autotune
        tracker.enable()
    else:
        tracker = TritonTrace.Tracker(
            output_dir, mode="offline", target="sm120")

    my_kernel[grid](A, B, C, M, N, K)  # capture or compile
    tracker.save_summary()
\end{minted}
\caption{Tracking a Triton Kernel for Simulation.}
\label{lst:triton-tracker}
\vspace{-8pt}
\end{figure}

Existing GPU simulators rely on hand-written CUDA microbenchmarks that bear little resemblance to the kernels running in production AI systems.
Evaluating simulator fidelity under realistic conditions requires running the same high-performance kernels that power real-world training and inference.
OpenAI Triton~\cite{triton} has emerged as a widely adopted framework for writing such kernels, striking a practical balance between performance and development productivity.
PyTorch's TorchInductor backend generates Triton code as its default GPU compilation path~\cite{torch-compile};
major LLM serving systems including vLLM~\cite{vllm} and SGLang~\cite{sglang} implement key compute kernels such as fused MoE in Triton;
and Liger Kernel~\cite{liger-kernel} provides a widely adopted Triton operator library for LLM training.
By supporting Triton workloads directly, FlashGPU-sim enables researchers to evaluate the simulator against production-level AI kernels, helping expose hardware bottlenecks while reducing interference from software inefficiencies.
This capability is critical for producing actionable hardware optimization insights.
Since Triton compiles to PTX, integration with FlashGPU-sim's PTX
interpretation model is feasible without modifying the compiler.

To bridge the gap between Triton's Python-level interface and FlashGPU-sim's
standalone execution interface, we develop a capture-and-replay framework.
By default, the framework operates online, capturing the compiled kernel,
launch configuration, arguments, and reference outputs during native GPU
execution, as illustrated in Figure~\ref{fig:triton-workflow}.
It packages the captured kernel and launch context into a standalone replay
harness, while the reference outputs enable automated correctness validation.
We first describe this online workflow and then present offline mode as an
optional path for preparing workloads without a physical GPU.

\myparagraph{Binary and Argument Capture}
During kernel loading, the framework captures the generated PTX and CUBIN
through Triton's \code{kernel\_load\_end} hook, together with kernel metadata
such as shared-memory usage and the number of warps.

Two additional hooks capture the arguments and results of each launch.
Before execution, the \code{launch\_enter} hook records scalar arguments,
serializes tensor arguments to binary files, and snapshots their contents.
Because Triton does not annotate output arguments, the
\code{launch\_exit} hook compares tensors against their pre-launch snapshots
and stores modified tensors as reference outputs.
Launch-specific harness generation assumes a one-to-one correspondence between
captured runtime arguments and PTX-level parameters, requiring Triton to preserve
all runtime arguments during compilation.

\myparagraph{Harness Generation}
At replay time, the generated CUDA harness loads a fatbinary assembled
from the captured PTX and CUBIN, restores the serialized tensor arguments and
recorded scalar values, allocates the required runtime scratch buffers, and
launches the kernel with the recorded grid dimensions, block dimensions, and
shared-memory configuration.
For online captures, it also compares the replayed outputs against the
recorded references.

\myparagraph{Online Mode}
The tracker observes native Triton execution through runtime hooks and
transparently handles Triton internals like runtime-injected
arguments and dynamic grid evaluation.
As shown in Figure~\ref{lst:triton-tracker}, tracking for an autotuned kernel
is enabled only after Triton selects and caches a configuration, and the
kernel is then re-invoked to capture the selected launch.

\myparagraph{Offline Mode}
To decouple Triton workload preparation from physical GPU availability,
offline mode hooks into Triton's JIT compilation path and redirects each
kernel invocation to \code{triton.compile}, bypassing native execution.
Users explicitly specify the target GPU architecture, and a single kernel
invocation generates the replay harness, as shown in
Figure~\ref{lst:triton-tracker}.
Since the kernel is not executed, offline mode cannot capture reference
outputs or perform the runtime profiling required for Triton autotuning.
An autotuned kernel therefore requires a single preselected
\code{triton.Config}.

\section{Validation}
\label{sec:validation}

\iflatexml
\else
\begin{table}[t]
\centering
\caption{MMA Peak Throughput.}
\label{tab:mma-peak}
\setlength{\tabcolsep}{3.5pt}
\footnotesize
\begin{tabular}{@{}llrrrrr@{}}
\toprule
\textbf{Type} & \textbf{Shape}
& \textbf{Peak}
& \multicolumn{2}{c}{\textbf{RTX~5090}}
& \multicolumn{2}{c}{\textbf{Simulator}} \\
\cmidrule(lr){4-5}\cmidrule(lr){6-7}
& &
& \textbf{Achieved} & \textbf{Eff.}
& \textbf{Achieved} & \textbf{Eff.} \\
\midrule
\multirow{2}{*}{FP16}
& M16N8K16 & 224.56 & 221.09 & 98.46\% & 205.38 & 91.46\% \\
& M16N8K8 & 224.56 & 110.99 & 49.43\% & 109.11 & 48.59\% \\
\midrule
BF16
& M16N8K8 & 224.56 & 110.97 & 49.42\% & 109.11 & 48.59\% \\
\midrule
\multirow{2}{*}{TF32}
& M16N8K8 & 112.33 & 110.56 & 98.42\% & 109.11 & 97.13\% \\
& M16N8K4 & 112.33 & 55.49 & 49.39\% & 54.56 & 48.57\% \\
\midrule
\multirow{2}{*}{INT8}
& M16N8K32 & 898.23 & 886.20 & 98.66\% & 764.12 & 85.07\% \\
& M16N8K16 & 898.23 & 439.81 & 48.96\% & 417.01 & 46.43\% \\
\bottomrule
\end{tabular}

\vspace{2pt}
\footnotesize
Peak and achieved throughput are in TOPS, normalized to 2580\,MHz.
\vspace{-4pt}
\end{table}

\fi

\iflatexml
\else
\begin{table}[t]
\centering
\caption{Distributed Shared Memory Validation.}
\label{tab:h200-dsm}
\setlength{\tabcolsep}{3.5pt}
\footnotesize
\begin{tabular}{@{}llrrr@{}}
\toprule
\textbf{Metric} & \textbf{Case}
& \textbf{H200} & \textbf{Simulator} & \textbf{Diff} \\
\midrule
\multicolumn{5}{l}{\textit{Distributed Shared Memory Latency (cycles)}} \\
\midrule
Remote load
& Mean latency
& 193.4 & 202.5 & $+4.7\%$ \\

Remote load
& Increment over local load
& 156.4 & 151.6 & $-3.0\%$ \\

SM-to-SM
& One-way latency
& 78.2 & 75.8 & $-3.0\%$ \\

Remote load
& Dependent round-trip
& 220.0 & 236.0 & $+7.2\%$ \\

Remote store
& Visibility latency
& 625.2 & 669.4 & $+7.1\%$ \\

Remote load
& Concurrent disjoint pairs
& 216.5 & 233.4 & $+7.8\%$ \\

\midrule
\multicolumn{5}{l}{\textit{Distributed Shared Memory Bandwidth (B/cycle)}} \\
\midrule

\multirow{2}{*}{Remote load}
& Unidirectional
& 19.88 & 17.73 & $-10.8\%$ \\
& Bidirectional
& 30.70 & 30.45 & $-0.8\%$ \\

\multirow{2}{*}{Remote store}
& Unidirectional
& 18.95 & 16.68 & $-11.9\%$ \\
& Bidirectional
& 36.31 & 35.79 & $-1.4\%$ \\

\multirow{2}{*}{TMA peer copy}
& Unidirectional
& 21.06 & 21.33 & $+1.3\%$ \\
& Bidirectional
& 41.37 & 42.50 & $+2.7\%$ \\

\multirow{2}{*}{Load + TMA}
& Co-direction
& 21.38 & 21.18 & $-0.9\%$ \\
& Counter-direction
& 28.80 & 29.53 & $+2.6\%$ \\

\midrule
\multicolumn{5}{l}{\textit{TMA Peer-Copy Bandwidth Scaling (B/cycle)}} \\
\midrule

\multirow{4}{*}{TMA peer copy}
& 2 SMs
& 41.1 & 42.6 & $+3.5\%$ \\
& 4 SMs
& 79.6 & 85.1 & $+6.9\%$ \\
& 8 SMs
& 163.9 & 169.9 & $+3.7\%$ \\
& 16 SMs
& 335.8 & 339.5 & $+1.1\%$ \\

\bottomrule
\end{tabular}
\vspace{-4pt}
\end{table}

\fi

\subsection{Methodology}
\label{subsec:validation-methodology}
We use the NVIDIA GeForce RTX~5090 as the primary platform for detailed validation, with 170~SMs, 32\,GB GDDR7 on a 512-bit bus (1792\,GB/s peak), and a 96\,MB L2 cache.
Additional experiments on H100, H200, and B200 further validate FlashGPU-sim across architectures.
Unless otherwise noted, the RTX~5090 SM and memory clocks are locked to 2580\,MHz and 14000\,MHz, respectively, to ensure reproducible cycle counts.
Hardware cycle counts are measured with \code{ncu} and compared against FlashGPU-sim's simulated cycles, with mean absolute percentage error (MAPE) serving as the primary accuracy metric.
The RTX~5090 AI workloads are implemented in Triton and captured via the extraction framework described in Section~\ref{subsec:triton}; tile sizes are determined by Triton's autotuner.
Simulation runs on an Intel Core i9-14900K with 64\,GB DDR5 RAM, 
while the multi-threaded performance experiments use an AMD EPYC 9115.

\subsection{Primitive Alignment}
\label{subsec:primitive-alignment}

\myparagraph{TMA Throughput}
The TMA bandwidth benchmark issues asynchronous bulk loads into a multi-stage shared-memory pipeline synchronized with \code{mbarrier}, while a single consumer warp performs minimal computation to keep the pipeline running.
The benchmark sweeps the number of active SMs from 1 to 170, tracing the full bandwidth curve from per-unit throughput to system-level saturation. 

\begin{figure}[t]
\centering
\iflatexml
\includegraphics[width=\textwidth]{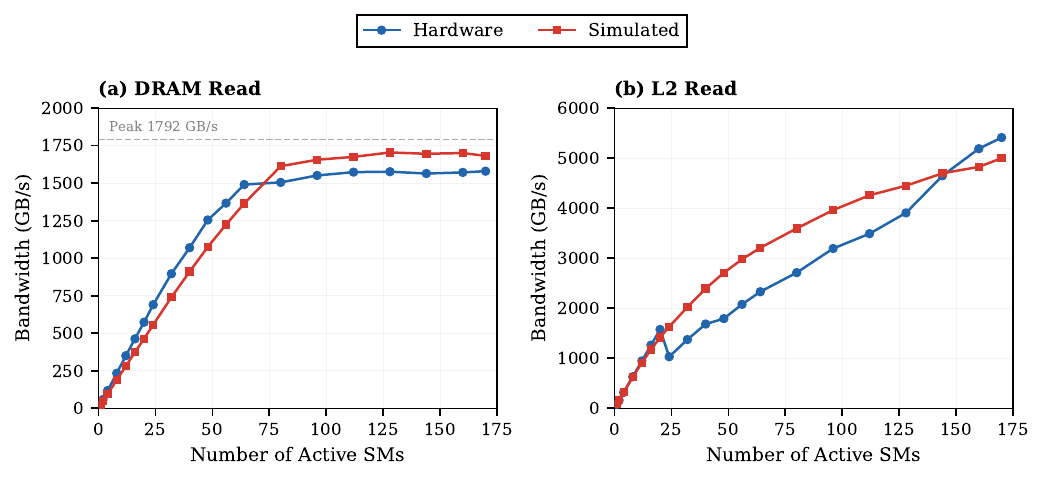}%
\else
\includegraphics[width=\columnwidth]{figs/tma-throughput.pdf}%
\fi
\caption{TMA Throughput Scaling.}
\label{fig:tma-throughput}
\vspace{-4pt}
\end{figure}

Figure~\ref{fig:tma-throughput} compares DRAM and L2 bandwidth scaling.
For DRAM bandwidth, hardware scales linearly at approximately 29~GB/s per SM until saturating near 80~SMs at 1580~GB/s.
The simulator follows the same saturation shape with 23.7~GB/s per SM in the linear region and saturates at 1682~GB/s, about 6\% above hardware.
For L2 bandwidth, single-SM throughput reaches 76\,GB/s on hardware and
81\,GB/s in simulation.
At full-chip scale, hardware and simulation reach 5.4\,TB/s and
5.0\,TB/s, respectively.
The primary discrepancy occurs between 20 and 80~SMs, where hardware
exhibits a transient plateau before resuming scaling.
Overall, the simulator captures the main DRAM and L2 scaling trends from
single-SM throughput to full-chip saturation.

\myparagraph{MMA Throughput}
The \code{mma.sync} throughput benchmark scales the issue-gap microbenchmark from Section~\ref{subsec:tensorcore} to all 170~SMs with sufficient ILP and blocks to saturate the Tensor Cores.
Table~\ref{tab:mma-peak} reports the achieved throughput.
As expected from the measured issue gap reported in Section~\ref{subsec:tensorcore}, smaller shapes reach only half the peak of their larger counterparts on both hardware and simulator.
For the full-throughput floating-point shapes, the simulator reaches 91--97\% of theoretical peak, close to the 98\% achieved on hardware.

\iflatexml
\begin{table}[t]
\centering
\caption{MMA Peak Throughput.}
\label{tab:mma-peak}
\setlength{\tabcolsep}{3.5pt}
\footnotesize
\begin{tabular}{@{}llrrrrr@{}}
\toprule
\textbf{Type} & \textbf{Shape}
& \textbf{Peak}
& \multicolumn{2}{c}{\textbf{RTX~5090}}
& \multicolumn{2}{c}{\textbf{Simulator}} \\
\cmidrule(lr){4-5}\cmidrule(lr){6-7}
& &
& \textbf{Achieved} & \textbf{Eff.}
& \textbf{Achieved} & \textbf{Eff.} \\
\midrule
\multirow{2}{*}{FP16}
& M16N8K16 & 224.56 & 221.09 & 98.46\% & 205.38 & 91.46\% \\
& M16N8K8 & 224.56 & 110.99 & 49.43\% & 109.11 & 48.59\% \\
\midrule
BF16
& M16N8K8 & 224.56 & 110.97 & 49.42\% & 109.11 & 48.59\% \\
\midrule
\multirow{2}{*}{TF32}
& M16N8K8 & 112.33 & 110.56 & 98.42\% & 109.11 & 97.13\% \\
& M16N8K4 & 112.33 & 55.49 & 49.39\% & 54.56 & 48.57\% \\
\midrule
\multirow{2}{*}{INT8}
& M16N8K32 & 898.23 & 886.20 & 98.66\% & 764.12 & 85.07\% \\
& M16N8K16 & 898.23 & 439.81 & 48.96\% & 417.01 & 46.43\% \\
\bottomrule
\end{tabular}

\vspace{2pt}
\footnotesize
Peak and achieved throughput are in TOPS, normalized to 2580\,MHz.
\vspace{-4pt}
\end{table}

\fi

\myparagraph{DSM Performance}
Table~\ref{tab:h200-dsm} compares FlashGPU-sim with independent H200 probes.
Remote shared-memory load latency, store visibility, and concurrent pair traffic show differences of 3.0--7.8\%.
Bidirectional and mixed DSM traffic remain within 10\%, 
while unidirectional remote load and store show slightly larger differences of about 11\%.
FlashGPU-sim also reproduces the near-linear scaling of TMA peer-copy bandwidth from 2 to 16 SMs, 
with differences below 4\% at three of four points and a maximum of 6.9\%.

\iflatexml
\begin{table}[t]
\centering
\caption{Distributed Shared Memory Validation.}
\label{tab:h200-dsm}
\setlength{\tabcolsep}{3.5pt}
\footnotesize
\begin{tabular}{@{}llrrr@{}}
\toprule
\textbf{Metric} & \textbf{Case}
& \textbf{H200} & \textbf{Simulator} & \textbf{Diff} \\
\midrule
\multicolumn{5}{l}{\textit{Distributed Shared Memory Latency (cycles)}} \\
\midrule
Remote load
& Mean latency
& 193.4 & 202.5 & $+4.7\%$ \\

Remote load
& Increment over local load
& 156.4 & 151.6 & $-3.0\%$ \\

SM-to-SM
& One-way latency
& 78.2 & 75.8 & $-3.0\%$ \\

Remote load
& Dependent round-trip
& 220.0 & 236.0 & $+7.2\%$ \\

Remote store
& Visibility latency
& 625.2 & 669.4 & $+7.1\%$ \\

Remote load
& Concurrent disjoint pairs
& 216.5 & 233.4 & $+7.8\%$ \\

\midrule
\multicolumn{5}{l}{\textit{Distributed Shared Memory Bandwidth (B/cycle)}} \\
\midrule

\multirow{2}{*}{Remote load}
& Unidirectional
& 19.88 & 17.73 & $-10.8\%$ \\
& Bidirectional
& 30.70 & 30.45 & $-0.8\%$ \\

\multirow{2}{*}{Remote store}
& Unidirectional
& 18.95 & 16.68 & $-11.9\%$ \\
& Bidirectional
& 36.31 & 35.79 & $-1.4\%$ \\

\multirow{2}{*}{TMA peer copy}
& Unidirectional
& 21.06 & 21.33 & $+1.3\%$ \\
& Bidirectional
& 41.37 & 42.50 & $+2.7\%$ \\

\multirow{2}{*}{Load + TMA}
& Co-direction
& 21.38 & 21.18 & $-0.9\%$ \\
& Counter-direction
& 28.80 & 29.53 & $+2.6\%$ \\

\midrule
\multicolumn{5}{l}{\textit{TMA Peer-Copy Bandwidth Scaling (B/cycle)}} \\
\midrule

\multirow{4}{*}{TMA peer copy}
& 2 SMs
& 41.1 & 42.6 & $+3.5\%$ \\
& 4 SMs
& 79.6 & 85.1 & $+6.9\%$ \\
& 8 SMs
& 163.9 & 169.9 & $+3.7\%$ \\
& 16 SMs
& 335.8 & 339.5 & $+1.1\%$ \\

\bottomrule
\end{tabular}
\vspace{-4pt}
\end{table}

\fi

\subsection{AI Workloads on RTX 5090}
\label{subsec:ai-workloads}

\begin{figure}[t]
\centering
\iflatexml
\includegraphics[width=\textwidth]{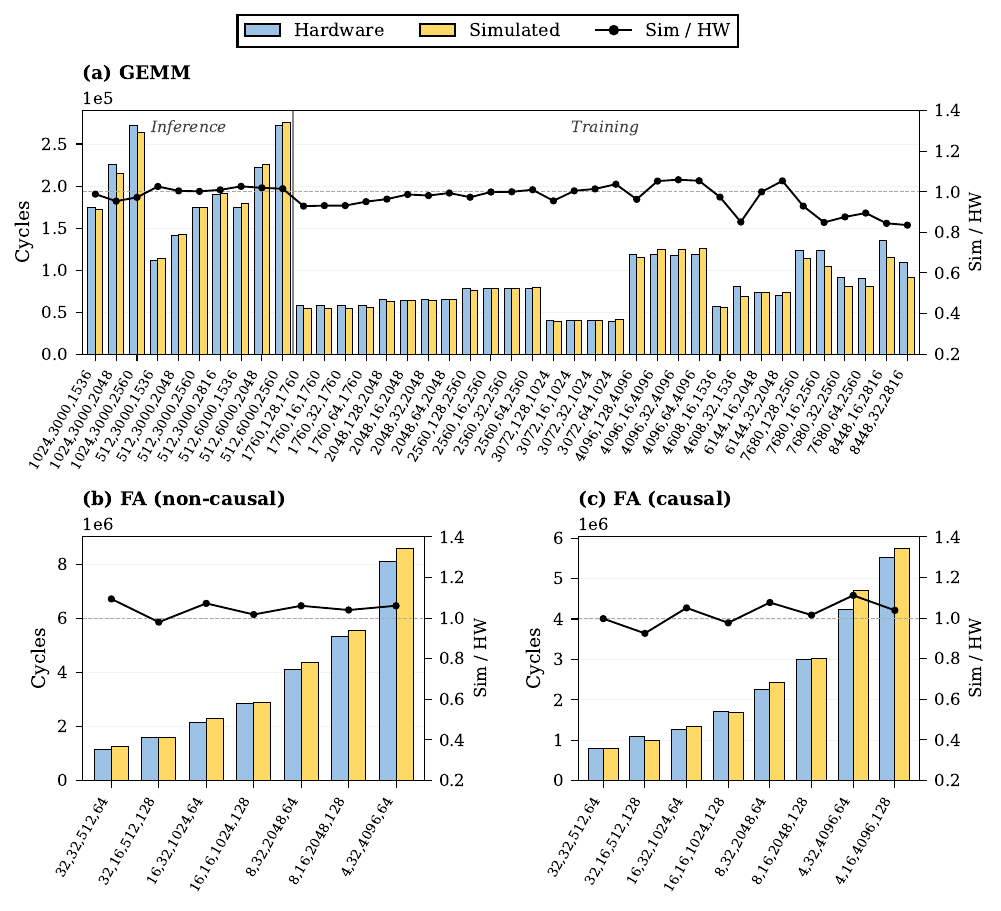}%
\else
\includegraphics[width=\columnwidth]{figs/kernel_validation.pdf}%
\fi
\caption{GEMM and FlashAttention Validation.}
\label{fig:kernel-validation}
\end{figure}

\myparagraph{Kernel-level Validation}
We first validate on GEMM and FlashAttention, two dominant kernels in modern LLM workloads.
The GEMM kernel is implemented in Triton using TMA for data movement and MMA for computation.
Matrix shapes are drawn from the inference-serving and training sets of DeepBench~\cite{deepbench}, yielding 10 inference and 30 training configurations.
The FlashAttention kernel is also implemented in Triton, using TMA for K/V block loads and MMA for the $QK^T$ and $PV$ computations.
Its 15 configurations are drawn from the FlashAttention-2~\cite{fa2} benchmark suite and cover two head dimensions ($d{=}64$ and $d{=}128$), sequence lengths from 512 to 4096, and both non-causal and causal modes.

Figure~\ref{fig:kernel-validation} and Table~\ref{tab:kernel-summary}
compare simulated and hardware cycles across all 55 kernel configurations
and summarize MAPE, bias, maximum diff, memory-traffic and occupancy
differences, and the fraction of configurations within 10\% diff.
Overall, 48 configurations fall within 10\% diff, including
34 of 40 GEMMs and 14 of 15 FlashAttention cases.
Memory traffic deviates by no more than 1.5\% across all categories,
and occupancy differences remain within 1.5\% except for GEMM inference
at 5.6\%.
GEMM achieves a MAPE of 4.6\% and FlashAttention 5.0\%,
yielding a combined MAPE of 4.8\%.

\begin{table}[t]
\centering
\caption{Kernel-level Validation Summary.}
\label{tab:kernel-summary}
\small
\resizebox{\columnwidth}{!}{%
\begin{tabular}{@{} l r r r r r r r @{}}
\toprule
\textbf{Category} & \textbf{\#} & \textbf{MAPE} & \textbf{Bias} & \textbf{Max} & \textbf{Traffic} & \textbf{Occ} & \textbf{Diff$<$10\%} \\
\midrule
Inference              & 10 &  1.9\% & $+$0.2\% &  4.6\% & $-$0.2\% & $+$5.6\% & 10/10 \\
\midrule
Training ($N{=}16$)    &  9 &  5.3\% & $-$4.0\% & 15.6\% & $-$1.3\% & $-$0.1\% &  7/9  \\
Training ($N{=}32$)    &  9 &  7.2\% & $-$4.4\% & 16.4\% & $-$1.3\% & $-$0.1\% &  6/9  \\
Training ($N{=}64$)    &  6 &  4.4\% & $-$1.0\% & 10.5\% & $-$1.5\% & $+$0.3\% &  5/6  \\
Training ($N{=}128$)   &  6 &  4.8\% & $-$4.8\% &  7.1\% & $-$0.9\% & $+$1.5\% &  6/6  \\
\cmidrule{1-8}
\textit{GEMM All (40)} & 40 & \textit{4.6\%} & \textit{$-$2.7\%} & \textit{16.4\%} & \textit{$-$1.0\%} & \textit{$+$1.6\%} & \textit{34/40} \\
\midrule
FA Non-causal          &  7 &  5.2\% & $+$4.6\% &  9.4\% & $-$0.2\% & $-$0.2\% &  7/7  \\
FA Causal              &  8 &  4.9\% & $+$2.4\% & 11.2\% & $-$0.4\% & $-$1.0\% &  7/8  \\
\cmidrule{1-8}
\textit{FA All (15)}   & 15 & \textit{5.0\%} & \textit{$+$3.4\%} & \textit{11.2\%} & \textit{$-$0.3\%} & \textit{$-$0.6\%} & \textit{14/15} \\
\bottomrule
\end{tabular}%
}
\end{table}

\begin{table}[t]
\centering
\caption{Llama3-8B layer-level inference validation.}
\label{tab:llama3-results}
\setlength{\tabcolsep}{3pt}
\small
\resizebox{\columnwidth}{!}{%
\begin{tabular}{@{} r l l r r r r @{}}
\toprule
\textbf{\#} & \textbf{Kernel} & \textbf{Function} & \textbf{Sim} & \textbf{NCU} & \textbf{Diff} & \textbf{Diff} \\
\midrule
\multicolumn{7}{@{}l}{\textit{Prefill: batch 2, sequence length 128}} \\
\midrule
 1 & RMSNorm & Input normalization &  62{,}794 &  62{,}044.7 & $+$749 & $+$1.2\% \\
 2 & Linear  & QKV projection & 319{,}315 & 332{,}203.6 & $-$12{,}889 & $-$3.9\% \\
 3 & GQA     & Causal prefill attention &  41{,}090 &  42{,}427.2 & $-$1{,}337 & $-$3.2\% \\
 4 & Linear  & Output projection + residual & 170{,}275 & 173{,}518.3 & $-$3{,}243 & $-$1.9\% \\
 5 & RMSNorm & FFN normalization &  62{,}794 &  63{,}653.1 & $-$859 & $-$1.3\% \\
 6 & Linear  & FFN up/gate projection & 940{,}773 & 988{,}824.6 & $-$48{,}052 & $-$4.9\% \\
 7 & SwiGLU  & SiLU + gate multiply &  23{,}672 &  25{,}995.3 & $-$2{,}323 & $-$8.9\% \\
 8 & Linear  & FFN down projection + residual & 537{,}878 & 557{,}109.7 & $-$19{,}232 & $-$3.5\% \\
\midrule
   & \multicolumn{2}{l}{\textit{Total}} & \textit{2{,}158{,}591} & \textit{2{,}245{,}776.3} & \textit{$-$87{,}185} & \textit{$-$3.9\%} \\
\midrule
\multicolumn{7}{@{}l}{\textit{Decode: batch 256, query length 1, KV length 128}} \\
\midrule
 1 & RMSNorm & Input normalization &  62{,}794 &  63{,}148.6 & $-$355 & $-$0.6\% \\
 2 & Linear  & QKV projection + KV-cache write & 320{,}897 & 333{,}182.4 & $-$12{,}285 & $-$3.7\% \\
 3 & GQA     & Single-token decode attention & 537{,}986 & 557{,}999.7 & $-$20{,}014 & $-$3.6\% \\
 4 & Linear  & Output projection + residual & 170{,}275 & 174{,}426.6 & $-$4{,}152 & $-$2.4\% \\
 5 & RMSNorm & FFN normalization &  62{,}794 &  61{,}592.0 & $+$1{,}202 & $+$2.0\% \\
 6 & Linear  & FFN up/gate projection & 940{,}773 & 987{,}916.0 & $-$47{,}143 & $-$4.8\% \\
 7 & SwiGLU  & SiLU + gate multiply &  23{,}672 &  25{,}512.5 & $-$1{,}840 & $-$7.2\% \\
 8 & Linear  & FFN down projection + residual & 537{,}878 & 557{,}018.0 & $-$19{,}140 & $-$3.4\% \\
\midrule
   & \multicolumn{2}{l}{\textit{Total}} & \textit{2{,}657{,}069} & \textit{2{,}760{,}795.7} & \textit{$-$103{,}727} & \textit{$-$3.8\%} \\
\bottomrule
\end{tabular}%
}
\vspace{-10pt}
\end{table}

\iflatexml
\else
\begin{figure*}[t]
\centering
\iflatexml
\includegraphics[width=\textwidth]{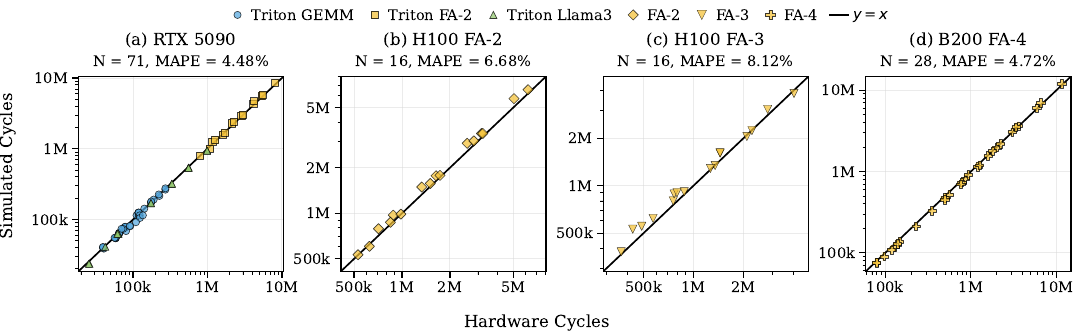}%
\else
\includegraphics[width=\textwidth]{figs/correlation.pdf}%
\fi
\caption{Overall Cycle Correlation.}
\label{fig:cross-arch-correlation}
\vspace{-8pt}
\end{figure*}

\fi

\myparagraph{LLM Inference}
To evaluate inference workloads beyond isolated kernels, we run the prefill and decode phases of Llama3-8B layers, covering the two primary execution stages of autoregressive serving.
This structure covers the operator mix typical of production LLM workloads: compute-bound QKV projections, output projections and SwiGLU MLPs, alongside memory-bound grouped-query attention (32 query heads, 8 KV heads) and residual connections. 
We adopt the standard Llama3-8B configuration: 4096 hidden channels, 128-dimensional heads, and a 14336-channel intermediate layer.
Residual additions are fused into the preceding matrix multiplications to reduce kernel-launch overhead.
The entire measured path is implemented in Triton and validated against PyTorch reference outputs for functional correctness.

The Llama3 layer workload involves alternating heterogeneous operators and frequent kernel launches, which can trigger frequency throttling at high SM clocks.
To reduce frequency variation during profiling, we lock the SM clock to 1.80~GHz, resulting in an \code{ncu}-reported average frequency of 1.77~GHz.
For hardware measurement stability, \code{ncu} profiling uses the
\code{--cache-control~all} and \code{--pipeline-boost-state~stable} flags.

Table~\ref{tab:llama3-results} reports per-kernel cycle validation results for Llama3-8B layer-level inference.
FlashGPU-sim remains accurate across the full layer execution in both prefill and decode.
The total cycle difference is $-3.9\%$ for prefill and $-3.8\%$ for decode,
with a MAPE of 3.5\% across the 16 kernel launches and a cycle-weighted MAPE of 3.9\%.

\subsection{Hopper and Blackwell}
\label{subsec:hopper-blackwell}

FlashGPU-sim also supports modern datacenter GPUs, including Hopper and
Blackwell.
We validate the support using the high-performance FlashAttention
implementations introduced in Section~\ref{sec:background}.
Specifically, we evaluate FlashAttention-2 (FA-2) and FlashAttention-3 (FA-3)
on H100, and extend the validation to FlashAttention-4 (FA-4) on B200.
These kernels exercise three distinct Tensor Core execution paths:
\code{mma.sync} with \code{cp.async} in FA-2, Hopper \code{wgmma} with TMA
in FA-3, and Blackwell \code{tcgen05} with TMEM and TMA in FA-4.
This complements the Triton-based validation above with official,
architecture-optimized FlashAttention implementations.

Figure~\ref{fig:cross-arch-correlation} compares simulated and hardware
cycles for the H100 and B200 FlashAttention experiments.
For reproducible profiling, the H100 and B200 SM clocks are set to
1.5\,GHz and 1.08\,GHz, respectively.
On H100, we evaluate 16 configurations each for FA-2 and FA-3,
which achieve MAPEs of 6.68\% and 8.12\%, respectively,
with 11 and 10 of 16 configurations showing differences below 10\%.
On B200, we evaluate 28 FA-4 configurations.
FA-4 achieves a MAPE of 4.72\%, with all 28 configurations showing differences below 10\%.

\iflatexml
\begin{figure*}[t]
\centering
\iflatexml
\includegraphics[width=\textwidth]{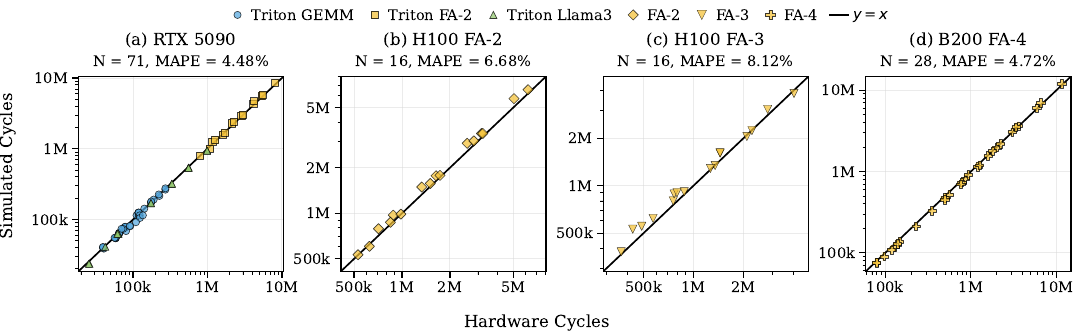}%
\else
\includegraphics[width=\textwidth]{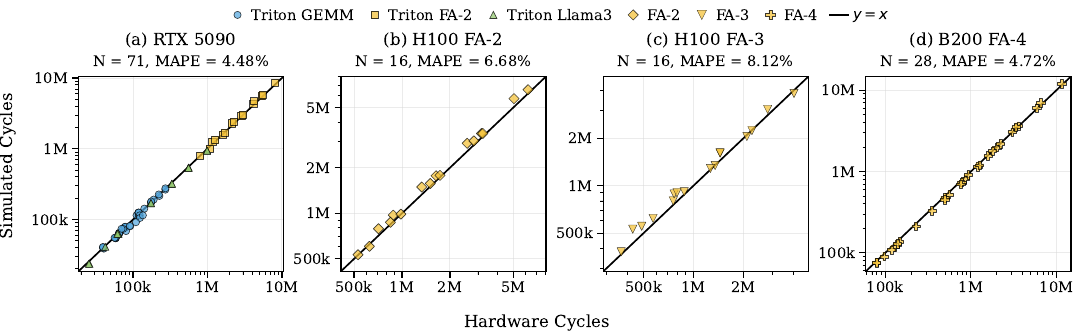}%
\fi
\caption{Overall Cycle Correlation.}
\label{fig:cross-arch-correlation}
\vspace{-8pt}
\end{figure*}

\fi

\subsection{Discussion}
\label{subsec:discussion}

\myparagraph{Validation Robustness}
Figure~\ref{fig:cross-arch-correlation} summarizes cycle-level correlation
across 131 workload configurations on RTX~5090, H100, and B200.
The evaluated kernels span more than two orders of magnitude in execution
cycles and include both Triton-generated workloads and architecture-optimized
FlashAttention implementations.
Across all four panels, simulated cycles closely track hardware, with
per-panel MAPEs ranging from 4.48\% to 8.12\%, showing consistent accuracy
across workload characteristics and GPU architectures.

\myparagraph{PTX-SASS Gap}
In controlled FA-4 experiments, two small configurations show cycle
differences above 40\% without reordering, which fall to within about
6\% when the reordering pass is enabled.
This highlights the performance impact of the PTX/SASS scheduling
mismatch discussed in Section~\ref{sec:simulator}.
A controlled ablation further shows that instruction-count reduction
alone does not explain this sensitivity: eliminating compiler-removed
register-pack operations reduces warp instructions by 8.84\%
but simulated cycles by only 2.19\%.
Together, these observations show that timing fidelity depends on
preserving the compiler-exposed dependency and scheduling structure,
rather than matching instruction count alone.

The impact of this mismatch also depends strongly on
latency-hiding parallelism.
In our RTX~5090 GEMM workloads, abundant resident warps and CTAs provide
enough thread-level parallelism to mask much of the timing distortion
from imperfect PTX ordering.
In contrast, highly optimized FlashAttention kernels expose less such
slack: FA-2 to FA-4 rely heavily on fine-grained asynchronous pipeline
overlap with relatively limited CTA-level concurrency.
In these kernels, local scheduling differences can delay synchronization
and producer-consumer handoff, thereby reducing the overlap between
computation and asynchronous data movement.

\iflatexml
\else
\begin{figure}[t]
\centering
\iflatexml
\includegraphics[width=\textwidth]{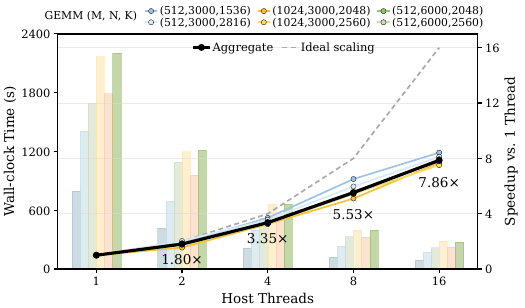}%
\else
\includegraphics[width=\columnwidth]{figs/sim-perf.pdf}%
\fi
\caption{Multi-threaded Simulation Speedup.}
\label{fig:sim-perf-scaling}
\vspace{-8pt}
\end{figure}

\fi

\subsection{Simulation Performance}
\label{subsec:simulation-performance}
We measure the wall-clock benefit of the multi-threaded execution
described in Section~\ref{subsec:sim-overview} on six inference-serving
GEMM workloads from Section~\ref{subsec:ai-workloads}, 
covering different CTA counts and per-SM workloads.
Each workload is run with multiple host-thread counts,
with each thread pinned to a distinct physical core to minimize scheduling interference.
Across repeated runs and thread counts, both functional outputs and
performance-simulation cycle counts remain stable.

Figure~\ref{fig:sim-perf-scaling} reports per-workload speedup normalized
to single-thread execution, together with the corresponding wall-clock time.
The aggregate speedup reaches $1.80\times$, $3.35\times$, $5.53\times$, and $7.86\times$
with 2, 4, 8, and 16 threads, respectively.
All six workloads follow the same trend, showing consistent scaling
across problem shapes as more physical cores are used.
At $7.86\times$ speedup, an hour-long single-threaded simulation is reduced
to about 7.6 minutes, making rapid design-space exploration and iterative
analysis practical.

\iflatexml
\begin{figure}[t]
\centering
\iflatexml
\includegraphics[width=\textwidth]{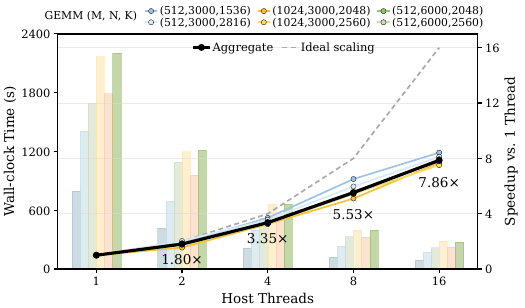}%
\else
\includegraphics[width=\columnwidth]{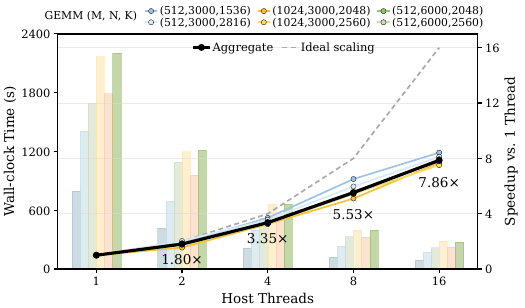}%
\fi
\caption{Multi-threaded Simulation Speedup.}
\label{fig:sim-perf-scaling}
\vspace{-8pt}
\end{figure}

\fi

\subsection{Case Study: Motivating Asynchronous Execution on H100}

\begin{figure}[t]
\centering
\iflatexml
\includegraphics[width=\textwidth]{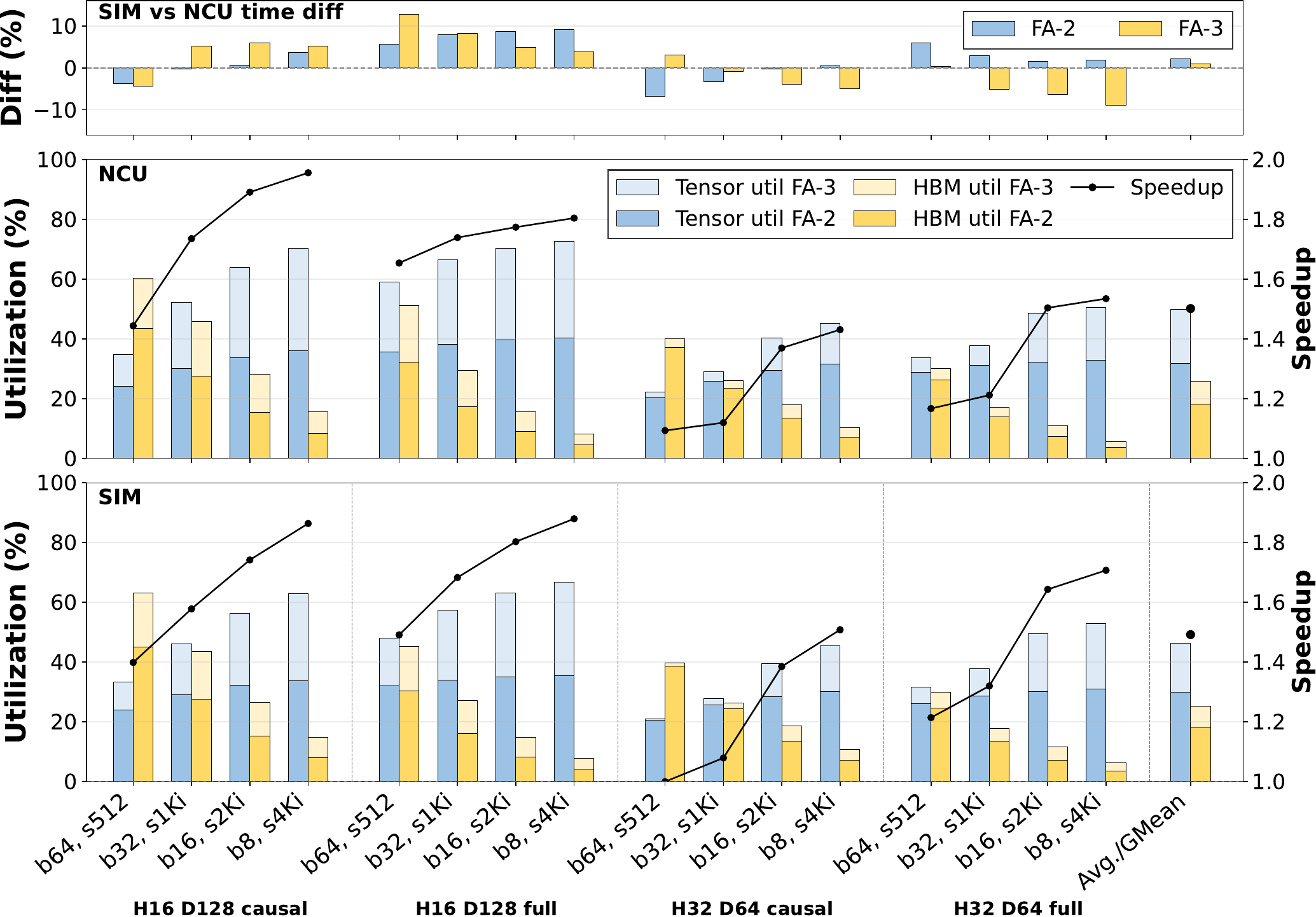}%
\else
\includegraphics[width=\columnwidth]{figs/H100-FA-sim-vs-ncu-crop.pdf}%
\fi
\caption{Simulating FA-2 vs. FA-3 on H100.}
\label{fig:h100-fa-sim}
\end{figure}

To further validate FlashGPU-sim's accuracy and ability to capture
architectural trade-offs, we reproduce the motivating FA-2 vs. FA-3 study on H100.
Building on the H100 calibration in
Section~\ref{subsec:hopper-blackwell}, we focus on small forward configurations
for detailed analysis and microarchitectural experiments.
Figure~\ref{fig:h100-fa-sim} compares tensor-core
utilization, HBM bandwidth, and end-to-end speedup, with the top row reporting
simulation error against ncu measurements.
FlashGPU-sim keeps average execution-time error below 5\% (max 13\%)
while preserving the cross-shape performance trend between FA-2 and FA-3.

To identify the root cause of FA-2's lower performance, we drill down into one
representative case in Table~\ref{tab:h100-fa-case}. Although FA-2 has
comparable occupancy, it relies on fine-grained synchronous
\texttt{mma.sync} and data-movement instructions\footnote{Although
\texttt{cp.async} is an asynchronous primitive, each instruction operates
at fine granularity and still requires per-transfer address generation and
control overhead compared to TMA. \texttt{mma.sync} is a synchronous MMA
instruction.}, which create substantial
front-end pressure:
20.97M \texttt{mma.sync}, 1.57M \texttt{cp.async}, and 6.55M
\texttt{ldmatrix} instructions in this case. Compared with FA-3's coarse
\texttt{wgmma}+TMA pipeline, this much larger instruction stream leads to
heavier queueing and waiting around MMA completion, and tensor units are less
effectively utilized when execution shifts to softmax and data movement. In
contrast, FA-3 is not evidently bound by MMA or bulk data movement in this
case; its dominant delay comes from scalar-dependency scoreboard pressure in
control and address-calculation paths, which indicates a more efficient tensor/data path
overall. A common optimization idea is to increase occupancy so that more warps
can provide ready instructions; however, even after we relax resource
constraints to raise FA-2 from 2 to 3 CTA/SM, runtime does not improve. This
confirms that higher occupancy does not translate to higher tensor
utilization when fine-grained dependence chains conflict with long MMA latency.

\begin{table}[tb!]
\centering
\caption{Microarchitectural comparison for one representative case
($B64$, $H16$, $S512$, $D128$, causal) on H100.}
\label{tab:h100-fa-case}

\newcommand{\rowrule}{%
  \noalign{\vskip 0pt}%
  \cmidrule(lr){1-4}
  \noalign{\vskip 0pt}%
}

\resizebox{\columnwidth}{!}{%
\begin{tabular}{lccc}
\toprule
Metric & FA-2 & FA-2-async & FA-3 \\
\midrule

Cycles / time
& \makecell{532,459\\354.973 us}
& \makecell{372,740\\248.493 us}
& \makecell{380,648\\253.765 us} \\
\rowrule

Sim occupancy
& 16.46\%
& 16.42\%
& 18.77\% \\
\rowrule

Tensor compute inst.
& \makecell{20.97M\\\texttt{mma.sync}}
& \makecell{20.97M\\\texttt{mma.sync}}
& \makecell{0.573M\\\texttt{wgmma}} \\
\rowrule

Data-move inst.
& \makecell{1.57M\\\texttt{cp.async}}
& \makecell{1.57M\\\texttt{cp.async}}
& \makecell{2.3K\\\texttt{tma/cp.async.bulk}} \\
\rowrule

Shared-matrix movement
& \makecell{6.55M\\\texttt{ldmatrix}}
& \makecell{6.55M\\\texttt{ldmatrix}}
& \makecell{0.262M\\\texttt{stmatrix}} \\
\rowrule

Total warp inst.
& 93.23M
& 93.23M
& 69.76M \\
\rowrule

Dominant stalls
& \makecell{MMAPipeThrottle\\(41.1\%)}
& \makecell{WaitTMA\\(19.1\%)}
& \makecell{Scalar-dep. scoreboard\\(40.7\%)} \\
\bottomrule
\end{tabular}}
\end{table}

We then run an improved experiment, FA-2-async, by inserting an
ideal queue between instruction issue and functional-unit execution for MMA and
data-movement operations to further decouple front-end issue from back-end
execution. FA-2-async reduces runtime from 532,459 cycles to 372,740 cycles
(about 30\%), nearly matching FA-3 (380,648 cycles) without changing peak
compute capability. This indicates that as matrix workloads scale,
coarse-grained decoupling and asynchronous execution become increasingly
important. Hopper advances this direction by exposing \texttt{wgmma} and TMA as
coarse-grained asynchronous primitives in both ISA and programming model,
improving hardware utilization by design. Overall, this case study
demonstrates that FlashGPU-sim can faithfully support microarchitecture-level
design exploration.

\section{Related Work}
\label{sec:related}

\myparagraph{GPU Simulators}
Publicly available GPU simulators span both execution-driven and trace-driven designs.
GPGPU-Sim~\cite{gpgpusim} remains the most widely adopted open-source NVIDIA simulator, providing cycle-accurate, execution-driven simulation by interpreting PTX on a modeled GPU pipeline.
Its architectural models, however, have not been updated beyond Volta, leaving it unable to represent the memory hierarchy and functional units of post-2020 NVIDIA designs.

Accel-Sim~\cite{accelsim} extended GPGPU-Sim with a trace-driven frontend that operates on NVIDIA's machine ISA, SASS.
Its workflow separates a tracing phase, which records the dynamic instruction stream on real hardware under NVBit instrumentation, from a replay phase, which feeds that fixed stream into a timing model derived from GPGPU-Sim.
This approach enabled support up to Ampere and made it possible to simulate closed-source libraries such as cuDNN without requiring source code.
More recently, Huerta~\etal~\cite{Micro2025DissectingGPUCore} proposed an improved timing model with compiler-assisted dependency tracking and multi-threaded simulation, also built on a trace-driven methodology.
More general full-system infrastructures such as gem5 and its later 20.0+ release are widely used in CPU and SoC research~\cite{gem5,gem5v20}; gem5-gpu extends that ecosystem with heterogeneous CPU-GPU modeling~\cite{gem5gpu}, but the public GPU support lacks modern NVIDIA asynchronous support.
On the AMD side, MGPUSim~\cite{mgpusim} targets GCN-era architectures and does not model NVIDIA-specific mechanisms.
Multi2Sim~\cite{multi2sim} supports both vendors but stops at Kepler.

\begin{figure}[tb!]
\centering
\iflatexml
\includegraphics[width=\textwidth]{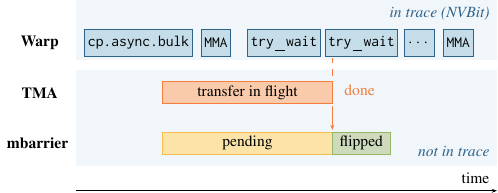}%
\else
\includegraphics{figs/trace_driven.pdf}%
\fi
\caption{Observability of Trace-driven Simulation.}
\label{fig:trace-async}
\vspace{-8pt}
\end{figure}

As illustrated in Figure~\ref{fig:trace-async}, trace-driven simulation separates \emph{what} a kernel executes from \emph{when} each operation completes: the tracer records every SASS instruction a warp issues, and the performance model replays this sequence to compute cycle-level timing.
This suffices for synchronous code, where the instruction stream is largely independent of hardware timing.
Asynchronous mechanisms, however, introduce hardware-side events that do not appear in the warp instruction stream.

When a bulk transfer completes, the copy engine updates memory state and signals the associated barrier; those events must be modeled internally rather than recovered from the trace.
Since the execution timing of these asynchronous events directly dictates when barriers release and control flow resolves, any architectural variation that alters timing can shift the resulting instruction stream.
As a result, a fixed trace preserves only the execution observed on the original hardware and may no longer match the execution induced by the modified architecture.
Execution-driven simulation overcomes this limitation by dynamically determining subsequent execution based on simulated timing and state.

\myparagraph{Adjacent AI Modeling Frameworks}
Several adjacent tools study AI systems at other abstraction levels rather than detailed GPU-core timing. ASTRA-SIM and ASTRA-sim2.0 model distributed training systems and communication hierarchies~\cite{astrasim,astrasim2}. Timeloop, MAESTRO, Accelergy, SCALE-Sim, and Aladdin focus on accelerator mapping, energy estimation, and design-space exploration~\cite{timeloop,maestro,accelergy,scalesim,aladdin}, while STONNE and NNASim provide accelerator-centric timing models for DNN inference hardware~\cite{stonne,nnasim}. Other emerging hardware domains have their own modeling ecosystems as well, including processing-in-memory simulators such as PIMulator-NN and PIMSIM-NN~\cite{pimulatornn,pimsimnn}, wafer-scale architecture co-exploration efforts such as WSC-LLM and Cerebras WSE studies~\cite{wscllm,cerebraswse}, and chiplet DSE work such as Gemini~\cite{geminichiplet}. These frameworks answer important adjacent design questions, but they do not substitute for a cycle-accurate simulator of modern NVIDIA GPU kernel execution.

\myparagraph{Modern AI Software Stacks and Kernels}
Modern high-performance AI kernels now come from both hand-written and compiler-assisted software paths. Compiler frameworks such as Triton, PyTorch~2, and TileLang make custom kernel generation more accessible~\cite{triton,torch-compile,tilelang}, while hand-optimized kernels such as FlashAttention-2 and FlashAttention-3 show how aggressively software adapts to new architectural features~\cite{fa2,fa3}. Recent kernel abstractions such as ThunderKittens and libraries such as Liger Kernel further expose warp-specialized, memory-hierarchy-aware programming styles to end users~\cite{thunderkittens,liger-kernel}.
At the system level, serving frameworks such as vLLM and SGLang shape which kernels dominate real deployments~\cite{vllm,sglang}.
This diversity of software paths makes PTX-level compatibility especially important for simulator usability. FlashGPU-sim's Triton frontend improves accessibility by avoiding manual kernel extraction, while PTX consumption keeps the simulator compatible with kernels emitted by these stacks as long as they lower to PTX. These works motivate the need for realistic modern workloads, but they do not provide a validated cycle-level simulator for post-Ampere NVIDIA GPUs.

\myparagraph{GPU Micro-architecture Reverse Engineering}
Huerta~\etal~\cite{Micro2025DissectingGPUCore} reverse-engineer the GPU core pipeline, focusing on issue logic, register files, and instruction scheduling, and integrate their findings into an Ampere-based GPGPU-Sim model.
Luo~\etal~\cite{DissectingNvidiaHopperArchitecture} provide a comprehensive microbenchmarking characterization of Hopper, covering memory hierarchy, tensor cores, and TMA.
These efforts are complementary to our goal: they improve understanding of modern hardware behavior, but do not by themselves provide a broadly capable modern simulator for future design exploration. Our work calibrates against newer hardware and translates such characterizations into a simulator validated on end-to-end AI workloads.

\section{Conclusion}
\label{sec:conclusion}
This paper presented FlashGPU-sim, an open-source, execution-driven,
cycle-accurate GPU simulator for modern GPU architectures and AI workloads.
Through detailed microarchitectural characterization, FlashGPU-sim models
asynchronous data movement, fine-grained synchronization, Tensor Core
execution, and distributed shared memory across recent NVIDIA architectures.
The Triton extraction front-end enables direct simulation of optimized AI
kernels, while multi-threaded execution substantially improves simulation
throughput while preserving deterministic simulation behavior.

Across RTX~5090, H100, H200, and B200, FlashGPU-sim maintains consistent
timing accuracy across primitive-level characterization and end-to-end
AI workloads.
The H100 FlashAttention case study further demonstrates FlashGPU-sim's
ability to identify microarchitectural bottlenecks and evaluate asynchronous
design trade-offs.
Ongoing efforts focus on integrating the gem5 memory subsystem to improve
memory-system modeling fidelity and extending FlashGPU-sim to multi-GPU
configurations with NVLink modeling for distributed and large-scale AI workloads.

\bibliographystyle{ref/IEEEtran}
\bibliography{ref/reference}

%
%

\let\AEOriginalTTDefault\ttdefault
\renewcommand{\ttdefault}{zi4}
\appendix[Artifact Appendix]
\label{sec:ae}
\setlength{\emergencystretch}{1em}

\subsection{Abstract}

The artifact packages FlashGPU-sim with the evaluated SM120 RTX~5090 configuration, CUDA microbenchmarks, Triton workloads, prepared hardware measurements, and self-contained traces. A unified dispatcher manages the reproduction scripts to regenerate the characterization and workload-simulation data, and rebuild \Cref{fig:tma-staircase,fig:tma-throughput,tab:mma-peak,fig:kernel-validation,fig:sim-perf-scaling} and \Cref{tab:mbarrier-wait,tab:mma-timing,tab:kernel-summary,tab:llama3-results}. An RTX~5090 GPU is preferred for collecting fresh measurements, while prepared measurements and traces provide a GPU-free fallback for the main simulation results.

\subsection{Artifact check-list}

{\small
\begin{itemize}[nosep,leftmargin=*]
  \item {\bf Compilation:} CUDA Toolkit 12.8, GCC/G++ with C++17 and OpenMP
  support, and GNU Make.
  \item {\bf Data set:} Included; enables optional GPU-free reproduction.
  \item {\bf Hardware:} An x86-64 host with at least 16 logical cores; 64\,GB
  RAM is recommended. Fresh GPU measurements require an RTX~5090 GPU.
  \item {\bf Run-time state:} GPU runs require \texttt{sudo} for clock control,
  and ncu profiling requires
  performance-counter access.
  \item {\bf Metrics:} Cycle counts and percentage differences.
  \item {\bf Experiments:} Primitive characterization of TMA completion,
  mbarrier polling, and MMA timing; TMA and MMA throughput alignment; GEMM,
  FlashAttention, and Llama3-8B workload validation; and simulator host-thread
  scaling.
  \item {\bf Disk space:} Approximately 40\,GB.
  \item {\bf Preparation time:} Approximately one hour.
  \item {\bf Experiment time:} Allow approximately 24 hours for the complete
  suite; detailed per-experiment runtime estimates are provided in the README.
  \item {\bf Publicly available?:} Yes.
  \item {\bf Archived?:} \url{https://doi.org/10.5281/zenodo.21537300}
\end{itemize}
}

\subsection{Description}

\subsubsection{How to access}

The artifact is archived on Zenodo at
\url{https://doi.org/10.5281/zenodo.21537300}. All commands are run from the
root of the extracted archive, with complete instructions provided in the
top-level \path{README.md}.

\subsubsection{Hardware dependencies}

An x86-64 host with at least 16 logical cores is required; 64\,GB RAM is
recommended. An RTX~5090 GPU is required for the TMA completion-time
staircase, mbarrier polling, and MMA timing microbenchmarks and for fresh
measurements of TMA/MMA throughput and the GEMM, FlashAttention, and Llama3-8B
workloads. Datasets for no-GPU reproduction of the throughput and AI-workload
simulations are also provided. Host-thread scaling is simulator-only.

\subsubsection{Software dependencies}

All workflows require Bash, GNU Make, GCC/G++ with C++17 and OpenMP support,
CUDA Toolkit 12.8, and Python 3.12. Python packages are pinned in
\path{common/requirements.txt}: PyTorch 2.9.0, Triton 3.5.0, NumPy 2.4.0, and
Matplotlib 3.10.8. Native execution additionally requires an NVIDIA driver
(version 580.82.07 was used for validation); profiling requires NVIDIA
Nsight Compute (ncu).

\subsubsection{Data sets}

\path{datasets/} contains RTX~5090 ncu reports for TMA throughput and the AI
workloads, an MMA throughput summary, and self-contained Triton traces for 40
GEMMs, 15 FlashAttention configurations, and 16 Llama3-8B layer launches.
These read-only inputs support optional GPU-free reproduction of the throughput
and AI-workload simulations; host-thread scaling reuses six of the GEMM traces.

\subsection{Installation}

From the artifact root, create the Python environment and build FlashGPU-sim
once. After compilation, start a new clean shell, return to the artifact root,
and export \texttt{CUDA\_INSTALL\_PATH} again before running the experiments.

\begin{lstlisting}[basicstyle=\small\ttfamily,columns=fullflexible,keepspaces=true,breaklines=true,frame=lines,framerule=0.4pt,framesep=3pt,showstringspaces=false,morecomment={[l]{\#}},commentstyle=\color{gray}]
$ export CUDA_INSTALL_PATH=/path/to/cuda-12.8
$ ./common/setup_env.sh  # Create Python environment
$ cd FlashGPU_sim
$ source setup_environment
$ make -j4  # Build the FlashGPU-sim simulator
\end{lstlisting}

\subsection{Experiment workflow}

The suite uses the following experiment identifiers:

{\setlength{\topsep}{0.75\baselineskip}
\begin{center}
\small
\begin{tabular*}{\columnwidth}{@{\extracolsep{\fill}}cll@{}}
\toprule
\textbf{ID} & \textbf{Experiment} & \textbf{Paper Result} \\
\midrule
E1 & TMA throughput       & Fig.~\ref{fig:tma-throughput} \\
E2 & MMA throughput       & Table~\ref{tab:mma-peak} \\
E3 & GEMM validation      & Fig.~\ref{fig:kernel-validation}, Table~\ref{tab:kernel-summary} \\
E4 & FlashAttention validation & Fig.~\ref{fig:kernel-validation}, Table~\ref{tab:kernel-summary} \\
E5 & Llama3-8B validation & Table~\ref{tab:llama3-results} \\
E6 & TMA completion time & Fig.~\ref{fig:tma-staircase} \\
E7 & Mbarrier polling    & Table~\ref{tab:mbarrier-wait} \\
E8 & MMA timing          & Table~\ref{tab:mma-timing} \\
E9 & Multi-thread speedup & Fig.~\ref{fig:sim-perf-scaling} \\
\bottomrule
\end{tabular*}
\end{center}}

To reproduce a paper result, run \texttt{./run figure N} or
\texttt{./run table N}. The dispatcher performs the target-specific workload
execution and postprocessing. Please run only one GPU experiment at a time, 
including native execution and Nsight Compute profiling, as the dispatcher does not enforce
cross-process GPU locks.

Use \texttt{./run list} to view all available
targets, execution modes, and current status. Native execution is selected by
default where applicable. Append \texttt{--no-gpu} to use the prepared
datasets; the dispatcher does not fall back to this mode when the required
hardware is unavailable.

\begin{lstlisting}[basicstyle=\small\ttfamily,columns=fullflexible,keepspaces=true,breaklines=true,frame=lines,framerule=0.4pt,framesep=3pt,showstringspaces=false,morecomment={[l]{\#}},commentstyle=\color{gray}]
$ ./run list  # Check all experiments and status
$ ./run figure 10  # Run with GPU
$ ./run figure 10 --no-gpu  # Use prepared data
$ ./run table VII --no-gpu
\end{lstlisting}

Prerequisites for derived targets are not run automatically. Complete them
manually in the same execution mode: run \Cref{fig:kernel-validation} before
\Cref{tab:kernel-summary}. Final paper artifacts
are published under \path{reproduce/}; workload logs and intermediate results
remain under each experiment's \path{results/<run-id>/}.

\subsection{Evaluation and expected results}

The overall evaluation of this artifact targets both workflow integrity and simulation accuracy. The expected key results are:

\begin{itemize}[nosep,leftmargin=*]
  \item \Cref{fig:tma-staircase,fig:tma-throughput,tab:mma-peak} and \Cref{tab:mbarrier-wait,tab:mma-timing} reproduce the TMA, mbarrier, and MMA primitive characterizations and throughput trends.
  \item \Cref{fig:kernel-validation} contains 40 GEMM and 15 FlashAttention points demonstrating an aggregate MAPE below 10\%; \Cref{tab:kernel-summary} reports their corresponding MAPE, bias, maximum error, traffic, and occupancy metrics.
  \item \Cref{tab:llama3-results} contains 16 Llama3 launches with a MAPE below 10\% for both prefill and decode.
  \item \Cref{fig:sim-perf-scaling} reproduces the scaling trend of six GEMMs at 1/2/4/8/16 host threads, while simulated cycles for each GEMM remain identical across all thread counts.
\end{itemize}

\subsection{Experiment customization}

\begin{itemize}[nosep,leftmargin=*]
  \item {\bf Execution flags:}
  \begin{itemize}[nosep,leftmargin=1.4em]
    \item \texttt{--no-gpu}: Use prepared datasets for E1--E5 reproduction.
    \item \texttt{--refresh}: Regenerate results instead of reusing them.
    \item \texttt{--dry-run}: Show requirements and commands without execution.
  \end{itemize}
  \item {\bf Staged execution and recovery:} Each E1--E9 experiment directory
  provides a \path{run.sh} script that, depending on the experiment, exposes
  \texttt{native}, \texttt{trace}, \texttt{ncu}, \texttt{sim},
  \texttt{summary}, and \texttt{plot} stages. Use \texttt{--run-dir} to resume
  an existing result directory. Detailed stage commands are provided in the
  corresponding READMEs.
  \item {\bf Partial Reproduction:} Run only the GEMM or
  FlashAttention part of \Cref{fig:kernel-validation} with
  \texttt{./run figure 10 gemm} or
  \texttt{./run figure 10 fa}. Each command
  publishes a standalone plot; the combined target becomes available after
  both workloads complete in the same mode.
\end{itemize}

\let\ttdefault\AEOriginalTTDefault

\end{document}